\documentclass[12pt]{article}
\usepackage[cp1251]{inputenc}
\usepackage[english]{babel}
\usepackage{amsmath,amssymb,hyperref}
\usepackage[table]{xcolor}

\newcommand{\fnm}{\footnotemark}
\newcommand{\fnt}{\footnotetext}

\renewcommand{\theequation}{\arabic{section}.\arabic{equation}}
\catcode`\@=11 \@addtoreset{equation}{section}\catcode`\@=12

\begin{document}

\begin{center}

  \large \bf
   Triple $M$-brane configurations and preserved supersymmetries
\end{center}

\vspace{15pt}

\begin{center}

  \normalsize\bf
        A.A. Golubtsova\fnm[1]\fnt[1]{siedhe@gmail.com}$^{, b,c}$
        and   V.D. Ivashchuk\fnm[2]\fnt[2]{ivashchuk@mail.ru}$^{, a, b}$

 \vspace{7pt}

 \it (a) \ \ \ Center for Gravitation and Fundamental
 Metrology,  VNIIMS, 46 Ozyornaya Str., Moscow 119361, Russia  \\

 (b) \  Institute of Gravitation and Cosmology,
 Peoples' Friendship University of Russia,
 6 Miklukho-Maklaya Str.,  Moscow 117198, Russia \\

(c) \  Laboratoire de Univers et Th\'{e}ories (LUTh), Observatoire de Paris,\\
Place Jules Janssen 5, 92190 Meudon, France \\

 \end{center}
 \vspace{15pt}

\begin{abstract}
We investigate all standard triple composite $M$-brane intersections defined on
products of Ricci-flat manifolds for preserving supersymmetries in eleven-dimensional $\mathcal{N} =1$ supergravity.
The explicit formulae for computing the  numbers of preserved supersymmetries are obtained,
which generalize the relations for topologically trivial flat factor spaces presented in the classification by Bergshoeff et al.
 We obtain certain examples of configurations preserving some fractions of supersymmetries, e.g. containing such factor spaces as $K3$,
$\mathbb{C}^{2}_{*}/Z_{2}$, a four-dimensional $pp$-wave manifold and the two-dimensional pseudo-Euclidean manifold $\mathbb{R}^{1,1}_{*}/Z_{2}$.
\end{abstract}

\section{Introduction}

The developments  of \cite{BL1, Gust,  ABJM} have led to a renewed interest in various aspects of supergravity.
Classical BPS configurations of intersecting branes play an essential role in studies of non-perturbative superstring and $M$-theories as well as
in establishing and proving new supergravity/gauge correspondences. Non-maximally supersymmetric solutions are important in many applications of superstring dualities.
Various intersections of $M$-branes in 11-dimensional supergravity \cite{CJS} provide a unified viewpoint, because a large class of solutions describing brane intersections
can be obtained by dimensional reduction and duality transformations.

In the basic $M2$- and $M5$-brane solutions, preserving half of the supersymmetries \cite{DS,Gu},
the worldvolumes  have been taken to be flat pseudo-Euclidean spaces $\mathbb{R}^{1,k}$  ($k = 2, 5$)
and the transverse spaces have been taken to be flat Euclidean ones $\mathbb{R}^{r}$  ($r = 8, 5$).
Nevertheless, the brane configurations defined on the spacetimes with more complicated geometry involved are of interest \cite{Smith,IMtop}.

First examples of $M2$-brane solutions, partially preserving the supersymmetry,  with  Ricci-flat 8-dimensional
transverse spaces and the flat brane worldvolumes  $\mathbb{R}^{1,2}$
have been obtained in \cite{DLPS} and \cite{GGPT}. The supersymmetric $M5$-brane solutions with the flat transverse spaces $\mathbb{R}^8$ and
Ricci-flat 6-dimensional brane worldvolumes have been found in \cite{BP}, \cite{K2} and \cite{FOF}. In \cite{IVD1} explicit formulae for fractional
numbers of  supersymmetries  for $M2$- and $M5$-brane solutions have been derived.

In this paper,  we study triple orthogonal intersections of composite $M$-branes defined on the manifold of the form
\begin{equation}\label{1.1}
M = M_{0} \times M_{1} \times \ldots \times M_{n},
\end{equation}
where all factor spaces $M_{i}$ are Ricci-flat manifolds. It should be noted that the study of
the flat case of the factor spaces was undertaken in a variety of works \cite{Ts1}-\cite{Gau}. In \cite{BREJS}
the classification of supersymmetric $M$-brane configurations on the product manifold with the factor spaces $M_{i} = \mathbb{R}^{k_{i}}$  was presented.
According to this work the amount of preserved supersymmetries is given by

\begin{equation}\label{1.2}
\mathcal{N} = 2^{-k} \quad \textrm{with} \quad k = 1,2,3,4,5.
\end{equation}

However, the relation (\ref{1.2}) is not well justified if $M$-brane configurations are taken into consideration on the product
of Ricci-flat manifolds (\ref{1.1}). In this case the fractional number of   supersymmetries $\mathcal{N}$ depends upon
several numbers of chiral parallel (i.e.  covariantly constant) spinors on certain factor spaces $M_{i}$ and brane
sign factors $c_{s}$, which define the orientations of brane worldvolumes.

For clarity, we remind that the metric for orthogonally intersecting $M$-branes is split into several parts:
the common worldvolume, the relative transverse space and the totally transverse one. The classification
of possible factor spaces contained in  worldvolumes and transverse spaces and
admitting parallel spinors can be given in terms of the holonomy groups, see \cite{BP},\cite{FOF},\cite{IVD2} and references therein.
While for intersections of two $M$-branes, as well as for the case of single $M2$- and $M5$-branes,
one can use the results obtained in works \cite{Wang},\cite{Baum},
finding non-trivial examples for triple $M$-brane configurations is complicated by increasing dominance
of low-dimensional flat manifolds (of dimensions 1, 2 and 3) among factor spaces.

The purpose of this work is to find out the relations for the amounts of preserved supersymmetries for
triple $M$-brane configurations defined on the product of Ricci-flat manifolds. The cases of one and two $M$-branes were considered in
\cite{IVD1} and  \cite{IVD2}, respectively. Here we deal with non-localized composite brane solutions with a vanishing contribution from the Chern-Simons term.
However this may be a starting point for future considering localized brane solutions, which are of interest in view of possible applications by using the gravity/gauge
correspondence \cite{MaM}.

The structure of the paper is as follows. In section 2 we present a set up, main definitions and notations.
Here we use Propositions 1 and 2 from the previous work \cite{IVD2} which reduce the solutions to generalized Killing equations
to a search of  parallel (i.e. covariantly constant) spinors on the product manifold (\ref{1.1}) obeying three algebraic equations.
These equations  depend upon a brane configuration and  brane sign factors.
In section 3 we find  relations for fractional numbers of preserved supersymmetries for triple $M$-brane solutions:  $M5 \cap M5 \cap M5$, $M2 \cap M2
\cap M5$ and  $M2 \cap M5 \cap M5$. For a completeness  we  start here by considering three electric branes $M2 \cap M2 \cap M2$ which was performed earlier in \cite{IVD2}.

\section{Generalized Killing spinor equations}

The bosonic action in $11$-dimensional supergravity is given by
\begin{equation}\label{2.1}
S_{act} = \int d^{11}z\sqrt{|g|}\left\{R[g]  - \frac{1}{2(4!)}F^{2} \right\} - \frac16 \int A\wedge F \wedge F,
\end{equation}
where
\begin{equation}\label{2.2}
F = dA = \frac{1}{4!}F_{NPQR}dz^{N}\wedge dz^{P} \wedge dz^{Q} \wedge dz^{R}
\end{equation}
is the 4-form field strength of the 3-form potential $A$.

The solutions to the equations of motion for the model (\ref{2.1}) are defined on the (oriented) warped product spin manifold of the form
\begin{equation}\label{2.3}
M = M_{0} \times M_{1} \times \ldots \times M_{n}
\end{equation}
with the metric
\begin{equation}\label{2.4}
g = e^{2\gamma(x)}g^{0} + \sum^{n}_{i=1}e^{2\phi^{i}(x)}g^{i},
\end{equation}
where $g^{0} = g^{0}_{\mu\nu}(x) d x^{\mu} \otimes d x^{\nu}$ is a
metric on the (oriented) spin manifold $M_{0}$ and $g^{i} =
g^{i}_{m_{i}n_{i}}(y_i)dy^{m_{i}}_{i}\otimes dy^{n_{i}}_{i}$ is a
metric on the (oriented) spin  manifold $M_{i}$, $i = 1,\ldots,n$. We denote
$d_{\nu} = \textrm{dim} M_{\nu}$, $\nu = 0,\ldots,n$;
$\displaystyle{\sum^{n}_{\nu = 0}d_{\nu}} = 11$.

The manifold (\ref{2.3}) allows a frame such that the metric $g =
g_{MN} dx^{M}\otimes dx^{N}$ ($M, N = 0, \ldots, 10 $) can be
represented in the following form

\begin{equation}\label{2.5}
g_{MN} = \eta_{AB}e^{A}_{\ \  M}e^{B}_{\ \ N}, \quad \textrm{with} \quad \eta_{AB} = \eta^{AB} = \eta_{A}\delta_{AB},
\end{equation}
where $e^{A} = e^{A}_{\ \ M}dx^{M}$ is the diagonalizing $11$-bein,
$\eta_{A} = \pm1$; $A,B = 0,\ldots, 10$. (Here $\eta_{A} = -1$ only for one value of  the index $A$.)

The required backgrounds must admit 32-component
 Majorana-spinors $\varepsilon$ such that the supersymmetry variation of the gravitino field $\delta\psi_{M}$ vanishes, i.e.
\begin{equation}\label{2.6}
(D_{M} + B_{M}) \varepsilon = 0.
\end{equation}
Here
\begin{equation}\label{2.7}
D_{M} = \partial_{M} + \frac{1}{4}\omega_{ABM}\hat{\Gamma}^{A}\hat{\Gamma}^{B}
\end{equation}
 is the spinorial covariant derivative, $\hat{\Gamma}^{A}$ are $32\times 32$ gamma matrices in an orthonormal frame satisfying the Clifford algebra relations
\begin{equation}\label{2.8}
\hat{\Gamma}_{A}\hat{\Gamma}_{B} + \hat{\Gamma}_{B}\hat{\Gamma}_{A} = 2 \eta_{AB}\mathbf{1}_{32},
\end{equation}
$\omega^{A}_{\ \ BM}$ is the spin connection and $\omega_{ABM} =
\eta_{AC}\omega^{C}_{\ \ BM} = - \omega_{BAM}$. (See also the
approach of Alekseevsky et al. \cite{Alekseevsky}.)

In (\ref{2.6}) $B_{M}$ is a matrix-valued covector field induced by the 4-form field strength $F$
\begin{equation}\label{2.9}
 B_{M} = \frac{1}{288}\left(\Gamma_{M}\Gamma^{N}\Gamma^{P}\Gamma^{Q}\Gamma^{R}
  - 12\delta^{N}_{M}\Gamma^{P} \Gamma^{Q} \Gamma^{R}\right)F_{NPQR},
\end{equation}
where $\Gamma_{M}$ are world gamma matrices obeying
\begin{equation}\label{2.10}
\Gamma_{M}\Gamma_{N} + \Gamma_{M}\Gamma_{N} = 2g_{MN}\mathbf{1}_{32}, \quad \textrm{with}\quad \Gamma_{M} = e^{A}_{\ \ M}\hat{\Gamma}_{A}.
\end{equation}
It should be noted that for any manifold $M_{l}$ with the metric $g^{l}$ one considers $k_{l} \times k_{l}$
$\Gamma$-matrices with $k_{l} = 2^{[d_{l}/2]}$ obeying
\begin{equation}\label{2.11}
\hat{\Gamma}^{a_{l}}_{(l)}\hat{\Gamma}^{b_{l}}_{(l)} + \hat{\Gamma}^{b_{l}}_{(l)}\hat{\Gamma}^{a_{l}}_{(l)} = 2\eta^{(l)a_{l}b_{l}}\mathbf{1}_{k_{l}}.
\end{equation}
Here we use alternative double-number notations for  indices:
$a_{j} = 1_{j},\ldots,(d_{j})_{j}$, where $d_{j}$ is the dimension of the manifold $M_{j}$, $j=0,\ldots,n$.

The local frame co-vectors are chosen in the following  form
\begin{equation}\label{2.11a}
(e^{A}_{\ \ M}) = \textrm{diag}\left(e^{\gamma}e^{(0)a}_{\ \ \ \ \mu}, e^{\phi^{1}}e^{(1)a_{1}}_{\ \ \ \ m_{1}},\ldots, e^{\phi^{n}}e^{(n)a_{n}}_{\ \ \ \ m_{n}}\right),
\end{equation}
where
\begin{equation}\label{2.11b}
g^{0}_{\mu\nu} = \eta^{(0)}_{ab}e^{(0)a}_{\ \ \ \ \mu}e^{(0)b}_{ \ \ \ \ \nu}, \quad g^{i}_{m_{i}n_{i}} = \eta^{(i)}_{a_{i}b_{i}}e^{(i)a_{i}}_{\ \ \ \ m_{i}}e^{(i)b_{i}}_{\ \ \ \ n_{i}},
\end{equation}
$i=1,\ldots,n$, and the signature matrix $(\eta_{AB})$ in (\ref{2.5}) can be written down in the components
\begin{equation}\label{2.11c}
\left(\eta_{AB}\right) = \textrm{diag}\left(\eta^{(0)}_{ab}, \eta^{(1)}_{a_{1}b_{1}}, \ldots, \eta^{(n)}_{a_{n}b_{n}}\right).
\end{equation}
In (\ref{2.11b})-(\ref{2.11c}) ($\eta^{(l)}_{a_{l}b_{l}}$) is a diagonal signature matrix for the metric $g^{l}$,
equipped by a set of (local) frame vectors with components $e^{(l)a_{l}}_{\ \ \ \ m_{l}}$, $l = 0,\ldots,n$. We put
\begin{equation}\label{2.11d}
\det{\left(e^{(l) a_{l}}_{\ \ \ \ m_{l}}\right)} > 0,
\end{equation}
$l = 0,\ldots,n$, i.e. for any $l$ the oriented set of $d_{l}$-beins $e^{(l)a_{l}}$ has the orientation compatible with the orientation of the manifold $M_{l}$.

Henceforth the following notation for the volume $d_{i}$-form on the manifolds $\left(M_{i},g^{i}\right)$ is used
\begin{equation}\label{2.5a}
\tau_{i} \equiv \sqrt{|g^{i}(y_{i})|}dy^{1}_{i}\wedge\ldots\wedge dy^{d_{i}}_{i}
\end{equation}
for $i = 1,\ldots,n$.

In this paper we continue our investigations of composite
$M$-brane solutions \cite{IMC} (with standard
intersection rules) defined on the product of $(n+1)$ Ricci-flat
manifolds $M_l$ and governed by several harmonic functions $H_s$ on $(M_0,g^0)$.

It was shown in \cite{IVD2} that the solutions to eqs. (\ref{2.6}),
corresponding to composite $M$-brane backgrounds, admit a representation in the following form
\begin{equation}\label{2.12}
\varepsilon = \left(\prod_{s\in S_{e}} H_{s}\right)^{-1/6}\left(\prod_{s\in S_{m}}H_{s}\right)^{-1/12}\eta,
\end{equation}
where  $\eta$ is a covariantly constant spinor on the spin manifold (\ref{2.3}) with the metric
\begin{equation} \label{2.4a}
\bar{g}= g^0 + \sum_{i=1}^{n} g^i,
\end{equation}

i.e.

\begin{equation} \label{2.13a}
  \bar{D}_{M} \eta = 0,
\end{equation}

satisfying projection conditions
\begin{equation}
\label{2.14} \hat{\Gamma}_{[s]}\eta = c_{s}\eta,
\end{equation}
$s \in S$, where the brane set $S$ contains two subsets $S_e$ and $S_m$ denoting electric and magnetic brane sets, respectively.
Under this convention in  (\ref{2.12}) harmonic functions $H_{s}$ with $s \in S_e$
correspond to electric $M2$-branes and $H_{s}$ with $s \in S_m$ correspond to magnetic $M5$-branes.

We remind that in (\ref{2.14}) $c_s = \pm 1$ are brane sign
factors and $\hat{\Gamma}_{[s]}$ are brane  operators, $s \in S$,
which are defined as follows:  for the electric case
\begin{equation}\label{2.14e}
          \hat{\Gamma}_{[s]} =  \hat{\Gamma}^{A_1}  \hat{\Gamma}^{A_2} \hat{\Gamma}^{A_3},
         \qquad  \ s \in S_e,
      \end{equation}
      where three indices describe the position of the $s$-th $M2$-brane
      worldvolume  and for the magnetic case
   \begin{equation}\label{2.14m}
      \hat{\Gamma}_{[s]} = \hat{\Gamma}^{B_1}  \hat{\Gamma}^{B_2} \hat{\Gamma}^{B_3}
      \hat{\Gamma}^{B_4}  \hat{\Gamma}^{B_5}, \qquad  \ s \in S_m,
    \end{equation}
      where five indices $B_1,B_2, B_3, B_4, B_5$ describe the position not occupied
      by the $s$-th $M5$-brane  worldvolume.
 These operators are idempotent, i.e. $(\hat{\Gamma}_{[s]})^2 =
 \mathbf{1}_{32}$ for all $s$.

 Eqs. (\ref{2.13a}) are equivalent to the following set of
 equations \cite{IVD2}

 \begin{equation}\label{2.13}
 \bar{D}_{m_{l}}\eta = 0,
 \end{equation}
 $l = 0, \ldots, n$.
Here
\begin{equation} \label{2.15}
\bar{D}_{m_{l}}  = \partial_{m_{l}} +
        \frac{1}{4}  \omega^{(l)}_{a_{l}b_{l}m_{l}} \hat{\Gamma}^{a_{l}} \hat{\Gamma}^{b_{l}},
\end{equation}
where $\omega^{(l)}_{a_l b_l m_l}  = \eta^{(l)}_{a_l c_l}
\omega^{(l) c_l}_{\ \ \ b_l m_l} $ and $\omega^{(l) a_l}_{\ \ \
b_l m_l}$ are components of the spin connection corresponding to
the metric $g^l$ equipped with diagonalizing $d_l$-bein vectors
$e^{(l)a_l}$,
$l = 0, ..., n$.\footnote{In \cite{IVD2} $\bar{D}_{m_{l}}$ was denoted by $\bar{D}^{(l)}_{m_{l}}$.}

In what follows  operators (\ref{2.15}) will generate the covariant spinorial derivatives corresponding the manifolds  $M_l$
\begin{equation}\label{0.38a}
D^{(l)}_{m_l} = \partial_{m_l} +  \frac{1}{4} \omega^{(l)}_{a_l b_l m_l}\hat{\Gamma}^{a_l}_{(l)} \hat{\Gamma}^{b_l}_{(l)},
\end{equation}
 $l = 0, ..., n$.

 In the next section we calculate the fractional numbers ${\cal N}$
 of unbroken supersymmetries (SUSY) for all triple $M$-brane configurations. Here we
 consider generic solutions to  Eqs. (\ref{2.6}) in  the
 form (\ref{2.12}). For any configuration we have
  \begin{equation}\label{1.4}
   {\cal N} = N/32,
  \end{equation}
 where $N$ is the dimension of  the linear space of solutions to
 algebraic equations (\ref{2.13a}) and (\ref{2.14}). (For special
 non-generic choices of harmonic functions $H_s$
 the real fractional numbers ${\cal N}$  of unbroken SUSY may be higher
 than those given by (\ref{1.4})).

{\bf Remark.} In this paper as in \cite{IVD2} we put  for
simplicity $\varepsilon (z) \in \mathbb{C}^{32}$. The imposing of the Majorana
condition will give the same number $N$ for the dimension of the
real linear   space of parallel Majorana spinors obeying
(\ref{2.13a}) and (\ref{2.14}).

\section{Triple $M$-brane configurations}

In this section we deduce relations for fractional numbers of supersymmetries preserved by
triple $M$-brane configurations defined on the product manifold (\ref{1.1}).

\subsection{$M2\cap M2 \cap M2$}

Let us consider the configuration of three electric 2-branes
intersecting over a point. The configuration  which is defined on
the manifold
\begin{equation}\label{5.1}
 M_{0} \times M_{1}  \times  M_{2} \times M_{3} \times M_{4},
 \end{equation}
with $d_0 = 4$, $d_1 = d_2 = d_3 = 2$ and  $d_4 = 1$,  can be presented as in Fig.1.

\begin{figure}[!h]
\begin{eqnarray}
g = \left\{
\begin{array}{ccccccccccccc}
- & - & - & - & \times &\times & - & - & - & - &\times & &:H_1 \nonumber \\
- & - & - & - & - &- & \times & \times & - & - &\times & &:H_2  \\
- & - & \hbox to 0pt{\hss$\underbrace{\hskip2.5cm}_{M_0}$\hskip 4mm\hss}- & -&-&\hbox to 0pt{\hss$\underbrace{\hskip 1cm}_{M_1}$\hskip 4mm\hss} -& -& \hbox to 0pt{\hss$\underbrace{\hskip 1cm}_{M_2}$\hskip 4mm\hss}- & \times &\hbox to 0pt{\hss$\underbrace{\hskip 1cm}_{M_3}$\hskip 4mm\hss} \times& \underbrace{\times}_{M_4} & &:H_3.
\end{array} \right.
\end{eqnarray}
 \caption{$M2\cap M2 \cap M2$-intersection over a point. $M_{0}$ is the totally transverse space, $M_{1}$,$M_{2}$,$M_{3}$ are the relative transverse spaces and $M_{4}$ is the common worldvolume.}
\end{figure}

For each figure we denote by $\times$ a coordinate corresponding to a worldvolume direction
and every direction transverse to the brane by $-$.

The solution is given by
 \begin{eqnarray}\label{5.2}
 g= H_1^{1/3} H_2^{1/3} H_3^{1/3}
 \{ g^0  + H_1^{-1} g^1 +  H_2^{-1} g^2
   +  H_3^{-1} g^3 + H_1^{-1} H_2^{-1} H_3^{-1} g^4 \},
 \\ \label{5.3}
 F =  c_1 d H^{-1}_1 \wedge  \tau_1 \wedge \tau_4 +
       c_2 d H^{-1}_2 \wedge  \tau_2 \wedge \tau_4
      +  c_3 d H^{-1}_3 \wedge  \tau_3 \wedge \tau_4                                             ,
 \end{eqnarray}
where $c^2_1 = c^2_2 = c^2_3 =1$; $H_1$, $H_2$, $H_3$ are harmonic
functions on $(M_0,g^0)$,   metrics $g^i$, $i = 0,1,2,3,4$, are
Ricci-flat (the last four metrics are flat). The metrics $g^i$, $i
= 0,1,2,3$, have Euclidean signatures and the metric $g^4$ has the
signature $(-)$. Here we put $M_4 = \mathbb{R}$, $g^4 = - dt \otimes dt$
($\tau_4 = dt$). The brane sets are $I_1 = \{1,4 \}$, $I_2 = \{2,4\}$ and $I_3 = \{3,4 \}$.

Using the rules of decomposition for $\Gamma$-matrices on product spaces from \cite{LPR} the set of $\Gamma$-matrices can be represented in the following form
 \begin{eqnarray}\label{5.4}
(\hat{\Gamma}^{A})= \begin{array}{ccccccccc}
   (\hat{\Gamma}^{a_{0}}_{(0)} & \otimes & \mathbf{1}_2 & \otimes & \mathbf{1}_2 & \otimes & \mathbf{1}_2 & \otimes & 1, \\
   \hat{\Gamma}_{(0)} & \otimes & \hat{\Gamma}_{(1)}^{a_{1}}& \otimes & \mathbf{1}_2 & \otimes & \mathbf{1}_{2}& \otimes & 1, \\
   i \hat{\Gamma}_{(0)} & \otimes & \hat{\Gamma}_{(1)} &\otimes & \hat{\Gamma}_{(2)}^{a_{2}} & \otimes &  \mathbf{1}_2 & \otimes & 1, \\
   \hat{\Gamma}_{(0)} & \otimes & \hat{\Gamma}_{(1)} & \otimes & \hat{\Gamma}_{(2)} & \otimes & \hat{\Gamma}_{(3)}^{a_{3}} & \otimes& 1, \\
   \hat{\Gamma}_{(0)} & \otimes & \hat{\Gamma}_{(1)} & \otimes & \hat{\Gamma}_{(2)} & \otimes & \hat{\Gamma}_{(3)} & \otimes & 1),
 \end{array}
  \end{eqnarray}
 where $\hat{\Gamma}^{a_{0}}_{(0)}$ are $4\times4$ gamma matrices corresponding to $M_{0}$:
 \begin{equation}\label{5.4d}
 \hat{\Gamma}^{a_{0}}_{(0)}\hat{\Gamma}^{b_{0}}_{(0)} + \hat{\Gamma}^{b_{0}}_{(0)}\hat{\Gamma}^{a_{0}}_{(0)} = 2 \delta_{a_{0}b_{0}}\textbf{1}_{4},
 \end{equation}
 $\hat{\Gamma}^{a_{i}}_{(i)}$ are $2\times2$ gamma matrices corresponding to $M_{i}$, $i = 1,2,3$:
\begin{equation}\label{5.4d}
\hat{\Gamma}^{a_{i}}_{(i)}\hat{\Gamma}^{b_{i}}_{(i)} + \hat{\Gamma}^{b_{i}}_{(i)}\hat{\Gamma}^{a_{i}}_{(i)} = 2 \delta_{a_{i}b_{i}}\textbf{1}_{2}.
 \end{equation}
 and
  \begin{equation}\label{5.4a}
 \hat{\Gamma}_{(0)} = \hat{\Gamma}_{(0)}^{1_0} \dots \hat{\Gamma}_{(0)}^{4_0}, \quad
 \hat{\Gamma}_{(i)} = \hat{\Gamma}_{(i)}^{1_i} \hat{\Gamma}_{(i)}^{2_i},
  \end{equation}
   obey
 \begin{equation}\label{5.4b}
 (\hat{\Gamma}_{(0)})^2 =  \mathbf{1}_4, \qquad  (\hat{\Gamma}_{(i)})^2 = - \mathbf{1}_2.
 \end{equation}

The spinor monomial can be written as follows
  \begin{equation}\label{5.6}
   \eta =  \eta_0(x) \otimes \eta_1(y_1)\otimes \eta_2(y_2)\otimes \eta_3(y_3)
   \otimes \eta_4(y_4),
   \end{equation}
  where $\eta_0 = \eta_0(x)$ is a $4$-component spinor  on $M_0$,
  $\eta_i = \eta_i(y_i)$ is a $2$-component spinor on $M_i$,
  $i = 1,2,3$,  and $\eta_4 = \eta_4(y_4)$ is a $1$-component spinor on   $M_4$.

    Due to   (\ref{5.4}) and (\ref{5.4b}) the covariant derivative acts on the spinor $\eta$ as follows
     \begin{eqnarray}\label{5.7a}
     \bar{D}_{m_0} \eta  =  (D_{m_0}^{(0)} \eta_0) \otimes
      \eta_1 \otimes \eta_2 \otimes  \eta_3 \otimes  \eta_4, \quad
     \bar{D}_{m_1} \eta =  \eta_0 \otimes (D_{m_1}^{(1)}
     \eta_1) \otimes \eta_2 \otimes  \eta_3 \otimes  \eta_4 , \\ \nonumber
     \bar{D}_{m_2} \eta =  \eta_0 \otimes \eta_1 \otimes (D_{m_2}^{(2)}
     \eta_2) \otimes  \eta_3 \otimes  \eta_4,  \quad
     \bar{D}_{m_3} \eta =  \eta_0 \otimes \eta_1 \otimes  \eta_2 \otimes (D_{m_3}^{(3)}
     \eta_3) \otimes  \eta_4, \\ \nonumber
     \bar{D}_{m_4} \eta = \eta_0 \otimes \eta_1 \otimes
     \eta_2 \otimes \eta_3  \otimes (D_{m_4}^{(4)}  \eta_4),
     \end{eqnarray}
   where $D^{(i)}_{m_i}$  is a covariant  derivative
   corresponding to  $M_i$, $i = 0,1,2,3$. Here $D^{(4)}_{m_4} = \partial_{m_4}$.

  The  operators corresponding to $M2$-branes are given by
   \begin{equation}\label{5.8a}
     \hat{\Gamma}_{[s]} =  \hat{\Gamma}^{1_1}  \hat{\Gamma}^{2_1} \hat{\Gamma}^{1_4} =
     - \hat{\Gamma}_{(0)}  \otimes   \mathbf{1}_2 \otimes \hat{\Gamma}_{(2)}
     \otimes \hat{\Gamma}_{(3)} \otimes 1
     \end{equation}
    for $s =  I_1$,
    \begin{equation}\label{5.8b}
     \hat{\Gamma}_{[s]} =  \hat{\Gamma}^{1_2} \hat{\Gamma}^{2_2} \hat{\Gamma}^{1_4}
     =   - \hat{\Gamma}_{(0)} \otimes \hat{\Gamma}_{(1)} \otimes \mathbf{1}_2
           \otimes \hat{\Gamma}_{(3)} \otimes 1
     \end{equation}
    for $s = I_2$ and
    \begin{equation}\label{5.8c}
     \hat{\Gamma}_{[s]} = \hat{\Gamma}^{1_3} \hat{\Gamma}^{2_3} \hat{\Gamma}^{1_4}
     =   - \hat{\Gamma}_{(0)} \otimes \hat{\Gamma}_{(1)} \otimes \hat{\Gamma}_{(2)}
            \otimes \mathbf{1}_2  \otimes 1
     \end{equation}
      for $s = I_3$.

   The supersymmetry constraints (\ref{2.14}) are satisfied if
     \begin{eqnarray}\label{5.8ca}
     \hat{\Gamma}_{(0)} \eta_0  = c_{(0)} \eta_0, \qquad  c_{(0)}^2 =  1,
      \\ \label{5.8cb}
     \hat{\Gamma}_{(j)} \eta_j  = c_{(j)} \eta_j, \qquad  c_{(j)}^2 = -1,
     \end{eqnarray}
   $j = 1,2,3$, and
    \begin{equation}\label{5.8d}
    - c_{(0)} c_{(2)} c_{(3)} = c_1, \qquad   - c_{(0)} c_{(1)} c_{(3)} = c_2, \qquad  - c_{(0)} c_{(1)} c_{(2)} = c_3.
     \end{equation}

 Then one obtains the following solution to SUSY equations (\ref{2.6})
 corresponding to the field configuration from (\ref{5.2}), (\ref{5.3})
 \begin{equation}\label{5.9}
 \varepsilon  = H_1^{-1/6} H_2^{-1/6} H_3^{-1/6}
 \eta_0(x) \otimes \eta_1(y_1) \otimes \eta_2(y_2) \otimes \eta_3(y_3) \otimes
 \eta_4.
 \end{equation}
 Here  $\eta_i$, $i = 0,1,2,3$, are chiral parallel spinors
 defined on $M_i$, respectively ($D^{(i)}_{m_i} \eta_i = 0$),
  obeying  (\ref{5.8ca}), (\ref{5.8cb}) and (\ref{5.8d}); $\eta_4$ is constant.

  Eqs. (\ref{5.8d}) have the following  solutions
  \begin{equation}\label{5.9aa}
   c_{(0)} = c_1 c_2 c_3, \qquad  c_{(j)}  = \pm  i c_j,
   \end{equation}
   $j = 1,2,3$.

  Thus, the number of linear independent solutions
  given by (\ref{5.9})-(\ref{5.9aa}) reads

  \begin{equation}\label{5.9a}
   N  = 32 {\cal N} =  n_0(c_1 c_2 c_3)
            \sum_{c = \pm 1} n_1( i c c_1) n_2( i c
            c_2) n_3(  i c c_3),
   \end{equation}
 where $n_j (c_{(j)})$ is the number of chiral parallel spinors on
 $M_j$, $j = 0,1,2,3$; see (\ref{5.8ca}) and (\ref{5.8cb}).

{\bf Examples.}

 Let  $M_1 = M_2 = M_3 = \mathbb{R}^2$. The manifold $\mathbb{R}^{2}$ has one parallel spinor
 of chirality $(+i)$ and one --- of chirality $(-i)$,
 hence all $n_j (ic) = 1$, $j = 1,2,3$, $c = \pm 1$, and one gets from (\ref{5.9a})
   \begin{equation}\label{5.11}
   {\cal N}  = \frac{1}{16} n_0( c_1 c_2 c_3).
  \end{equation}

  It is worth noting that the chirality of the spinor $\eta_{0}$ on the manifold $M_{0}$ is defined by
  the product of the brane sign constants $c_{1}$, $c_{2}$, $c_{3}$.
  \begin{description}
    \item[\{a\}]  For  $M_0 = \mathbb{R}^4$ we have $n_0(c) = 2$ and hence $\mathcal{N} = 1/8$ for all values  of $c_i$, $i = 1,2,3$.
    \item[\{b\}]  Consider the case of the curved transverse space. Let $M_{0} = K3$, $K3=CY_{2}$,
    which is a 4-dimensional Ricci-flat K\"{a}hler manifold with the holonomy group $SU(2) = Sp(1)$.
    The $K3$ surface has two parallel spinors of the same chirality. We put $n_0(1) = 2$ and $n_0(-1) = 0$.
    Then we get $\mathcal{N} = 1/8$ if  $c_1 c_2 c_3 = 1$  and ${\cal N } = 0$ if  $c_1 c_2 c_3 = - 1$.
    \item[\{c\}] One obtains the same result  for the conic space  $M_{0} = \mathbb{C}^{2}_{*}/Z_{2}$,
     where $\mathbb{C}^{2}_{*} =  \mathbb{C}^{2} \setminus  \{ {\bf 0 } \} $. It also has parallel spinors of the same chirality:
     the pair $(n_0(1), n_0(-1))$ is either $(2,0)$ or $(0,2)$ depending on the choice of the spin structure.
     The completion of $\mathbb{C}^{2}_{*}/Z_{2}$ is the orbifold $\mathbb{C}^{2}/Z_{2}$.
  \end{description}

\subsection{$M5\cap M5 \cap M5$}

According to the classification of $M$-brane configurations which is presented in \cite{BREJS}
there are three possible intersections of three magnetic branes depending on the position of the branes in the bulk space.

\textbf{(i)}

The first case of the solution describing three intersecting $M5$-branes  is defined on the manifold of the form
\begin{equation}\label{6.1}
M_{0}\times M_{1}\times M_{2}\times M_{3}\times M_{4},
\end{equation}
where $d_{0} = 1$, $d_{1}= d_{2}=d_{3}= 2$, $d_{4}=4$.
The configuration is given in Figure 2.

\begin{figure}[!h]
\begin{eqnarray}
g = \left\{
\begin{array}{ccccccccccccc}
- & \times & \times & - &  - & - & - & \times & \times & \times &\times & &:H_1 \nonumber \\
- & - & - & \times & \times &- & - & \times & \times & \times &\times & &:H_2  \\
\underbrace{-}_{M_{0}} & - & \hbox to 0pt{\hss$\underbrace{\hskip1cm}_{M_1}$\hskip 4mm\hss}- & -&\hbox to 0pt{\hss$\underbrace{\hskip 1cm}_{M_2}$\hskip 4mm\hss}-&\times &\hbox to 0pt{\hss$\underbrace{\hskip 1cm}_{M_3}$\hskip 4mm\hss} \times& \times & \times & \times & \hbox to 0pt{\hss$\underbrace{\hskip2.4cm}_{M_4}$\hskip 1.8cm\hss}\times  & &:H_3.
\end{array} \right.
\end{eqnarray}
 \caption{$M5\cap M5 \cap M5$-intersection over a 3-brane. $M_{0}$ is the totally transverse space, $M_{1}$, $M_{2}$, $M_{3}$ are the relative transverse spaces and $M_{4}$ is the common worldvolume. }
\end{figure}

The solution reads
\begin{equation}\label{6.2}
g = H^{2/3}_{1}H^{2/3}_{2}H^{2/3}_{3}\Bigl\{g^{0} + H^{-1}_{1}g^{1}  +  H^{-1}_{2}g^{2}+ H^{-1}_{3}g^{3} +
 H^{-1}_{1}H^{-1}_{2}H^{-1}_{3}g^{4}\Bigr\},
\end{equation}
\begin{equation}\label{6.2A}
F = c_{1}(\ast_{0}dH_{1})\wedge \tau_{2}\wedge \tau_{3} +
 c_{2}(\ast_{0}dH_{2})\wedge \tau_{1}\wedge \tau_{3} +
c_{3}(\ast_{0} dH_{3})\wedge \tau_{1}\wedge \tau_{2},
\end{equation}
where $c^{2}_{1} = c^{2}_{2} = c^{2}_{3} = 1$; $H_{1}$, $H_{2}$, $H_{3}$ are harmonic functions on $(M_{0},g^{0})$.
The metrics $g^{i}$, $i=0,1,2,3$, have Euclidean signatures and
the metric $g^{4}$ has the signature $(-,+,+,+)$.
The branes sets are $I_{1} = \{1,4\}$, $I_{2} = \{2,4\}$, $I_{3} = \{3,4\}$. Here and in what follows $\ast_{0} $ is the Hodge operator on $(M_0,g^0)$.

Using the rules of decomposition from \cite{LPR} one can write $\Gamma$-matrices in the following form
\begin{eqnarray}\label{6.3}
(\hat{\Gamma}^{A})=
\begin{array}{ccccccccc}
  (1 & \otimes & \hat{\Gamma}_{(1)} & \otimes & \hat{\Gamma}_{(2)} & \otimes & \hat{\Gamma}_{(3)} & \otimes & \hat{\Gamma}_{(4)}, \\
  1 & \otimes &  i\hat{\Gamma}^{a_{1}}_{(1)} & \otimes &  \hat{\Gamma}_{(2)} & \otimes &  \hat{\Gamma}_{(3)} & \otimes &  \hat{\Gamma}_{(4)}, \\
  1 & \otimes &  \mathbf{1}_{2} & \otimes &  \hat{\Gamma}^{a_{2}}_{(2)} & \otimes & \hat{\Gamma}_{(3)} & \otimes &  \hat{\Gamma}_{(4)}, \\
  1 & \otimes &  \mathbf{1}_{2} & \otimes &  \mathbf{1}_{2} & \otimes & i\hat{\Gamma}^{a_{3}}_{(3)} & \otimes &  \hat{\Gamma}_{(4)}, \\
  1 & \otimes &  \mathbf{1}_{2} & \otimes & \mathbf{1}_{2} & \otimes & \mathbf{1}_{2} & \otimes &  \hat{\Gamma}^{a_{4}}_{(4)}) \\
\end{array}
\end{eqnarray}
Here the operators $\hat{\Gamma}_{(i)}$, $i = 1,2,3,4$, are given by
\begin{eqnarray}\label{6.4}
\hat{\Gamma}_{(1)} = \hat{\Gamma}^{1_{1}}_{(1)}\hat{\Gamma}^{2_{1}}_{(1)}, \quad \hat{\Gamma}_{(2)} = \hat{\Gamma}^{1_{2}}_{(2)}\hat{\Gamma}^{2_{2}}_{(2)}, \quad
\hat{\Gamma}_{(3)} = \hat{\Gamma}^{1_{3}}_{(3)}\hat{\Gamma}^{2_{3}}_{(3)},
\quad \hat{\Gamma}_{(4)} = \hat{\Gamma}^{1_{4}}_{(4)}\hat{\Gamma}^{2_{4}}_{(4)}\hat{\Gamma}^{3_{4}}_{(4)}\hat{\Gamma}^{4_{4}}_{(4)}
\end{eqnarray}
obey
\begin{equation}\label{6.5}
(\hat{\Gamma}_{(i)})^{2}  = -\mathbf{1}_{2}, \quad
(\hat{\Gamma}_{(4)})^{2} = -\mathbf{1}_{4}
\end{equation}
with $i=1,2,3$.
$\hat{\Gamma}^{a_{1}}_{(1)}$,$\hat{\Gamma}^{a_{2}}_{(2)}$,$\hat{\Gamma}^{a_{3}}_{(3)}$ are $2\times 2$ $\Gamma$-matrices corresponding to $M_{1}$, $M_{2}$, $M_{3}$ respectively, $\hat{\Gamma}^{a_{4}}_{(4)}$ is a set of gamma matrices corresponding to $M_{4}$.

The covariant derivatives can be written down as
\begin{eqnarray}\label{6.6}
\bar{D}_{m_{1}} = \partial_{m_{1}} + \frac 14 \omega^{(1)}_{a_{1}b_{1}m_{1}}\left(1\otimes \hat{\Gamma}^{a_{1}}_{(1)}\hat{\Gamma}^{b_{1}}_{(1)}\otimes \mathbf{1}_{2}\otimes \mathbf{1}_{2}\otimes \mathbf{1}_{4}\right), \nonumber \\
\bar{D}_{m_{2}} = \partial_{m_{2}} + \frac 14
\omega ^{(2)}_{a_{2}b_{2}m_{2}}
\left(1\otimes \textbf{1}_{2}\otimes \hat{\Gamma}^{a_{2}}_{(2)}\hat{\Gamma}^{b_{2}}_{(2)}\otimes \mathbf{1}_{2}\otimes\mathbf{1}_{4}\right), \nonumber \\
\bar{D}_{m_{3}} = \partial_{m_{3}} +
\frac 14 \omega ^{(3)}_{a_{3}b_{3}m_{3}}
\left(1\otimes \mathbf{1}_{2}\otimes \mathbf{1}_{2}\otimes \hat{\Gamma}^{a_{3}}_{(3)}\hat{\Gamma}^{b_{3}}_{(3)}\otimes \mathbf{1}_{4}\right), \nonumber \\
\bar{D}_{m_{4}} = \partial_{m_{4}} + \frac 14
 \omega^{(4)}_{a_{4}b_{4}m_{4}}\left(1\otimes
\mathbf{1}_{2}\otimes \mathbf{1}_{2}\otimes \mathbf{1}_{2} \otimes \hat{\Gamma}^{a_{4}}_{(4)}\hat{\Gamma}^{b_{4}}_{(4)}\right),
\end{eqnarray}
where $\omega ^{(i) a_{i}}_{\ \ \ \ b_{i}c_{i}}$ are components of the
  spin connection corresponding to the manifold $M_{i}$,
$D^{(i)}_{m_{i}}$ is a covariant derivatives corresponding to $M_{i}$, $i=1,2,3,4$, $\bar{D}_{m_{0}} = \partial_{m_{0}}$
and $D^{(0)}_{m_{0}}= \partial_{m_{0}}$.

Let $\eta$ be represented in the following form
\begin{equation}\label{6.7}
\eta = \eta_{0}(x)\otimes \eta_{1}(y_{1})\otimes\eta_{2}(y_{2})\otimes \eta_{3}(y_{3})\otimes\eta_{4}(y_{4}),
\end{equation}
where $\eta_{0}(x)$ is a 1-component spinor on $M_{0}$, $\eta_{i} = \eta_{i}(y_{i})$ is a 2-component spinor on $M_{i}$, $i=1,2,3$,
 $\eta_{4} = \eta_{4}(y_{4})$ is a 4-component spinor on $M_{4}$.

Then for spinorial covariant derivatives we get relations (\ref{5.7a}).

The operators  corresponding to the $M5$-branes read
\begin{equation}\label{6.9}
\hat{\Gamma}_{[s]}= \hat{\Gamma}^{1_{0}}\hat{\Gamma}^{1_{2}}\hat{\Gamma}^{2_{2}}\hat{\Gamma}^{1_{3}}\hat{\Gamma}^{2_{3}} = 1 \otimes \hat{\Gamma}_{(1)}\otimes \mathbf{1}_{2}\otimes \mathbf{1}_{2}\otimes \hat{\Gamma}_{(4)}
\end{equation}
for $s = I_{1}$,
\begin{equation}\label{6.10}
\hat{\Gamma}_{[s]} =\hat{\Gamma}^{1_{0}}\hat{\Gamma}^{1_{1}}\hat{\Gamma}^{2_{1}}\hat{\Gamma}^{1_{3}}\hat{\Gamma}^{2_{3}} = 1 \otimes  \mathbf{1}_{2} \otimes \hat{\Gamma}_{(2)}\otimes \mathbf{1}_{2} \otimes \hat{\Gamma}_{(4)}
\end{equation}
for $s = I_{2}$ and
\begin{equation}\label{6.11}
\hat{\Gamma}_{[s]} =\hat{\Gamma}^{1_{0}}\hat{\Gamma}^{1_{1}}\hat{\Gamma}^{2_{1}}\hat{\Gamma}^{1_{2}}\hat{\Gamma}^{2_{2}} = 1\otimes \mathbf{1}_{2}\otimes \mathbf{1}_{2} \otimes \hat{\Gamma}_{(3)}\otimes \hat{\Gamma}_{(4)}
\end{equation}
for $s = I_{3}$.

Thus the spinor $\eta$ obeys the projection conditions (\ref{2.14}) if
\begin{equation}\label{6.12}
\hat{\Gamma}_{(j)}\eta_{j} = c_{(j)} \eta_{j}, \qquad c^{2}_{(j)} = -1,
\end{equation}
$\quad j =1,2,3,4$, and
\begin{equation}\label{6.13}
c_{(1)}c_{(4)} = c_{1}, \quad c_{(2)}c_{(4)} = c_{2}, \quad c_{(3)}c_{(4)} = c_{3}.
\end{equation}
The solution to the generalized Killing equations corresponding to the field configuration from (\ref{6.2}), (\ref{6.2A}) can be represented in the following form
\begin{equation}\label{6.14}
\varepsilon = \left(\prod^{3}_{s=1} H^{-\frac{1}{12}}_{s}\right) \eta_{0} \otimes \eta_{1}(y_{1}) \otimes \eta_{2}(y_{2})\otimes\eta_{2}(y_{3})\otimes\eta_{4}(y_{4}),
\end{equation}
where $\eta_{i}$, $i=1,2,3,4$, are parallel spinors defined on $M_{i}$, respectively $\left(D^{(i)}_{m_{i}}\eta_{i} = 0\right)$, obeying (\ref{6.12}) and (\ref{6.13}), and
$\eta_{0}$ is a constant  one-dimensional spinor on $M_{0}$.

Eqs. (\ref{6.13}) have the following solutions
\begin{equation}\label{6.15}
c_{(1)} = -ic_{1}, \quad c_{(2)} = -ic_{2}, \quad c_{(3)} = - i c_{3}, \quad c_{(4)} =  i,
\end{equation}
or
\begin{equation}\label{6.16}
c_{(1)} = i c_{1}, \quad c_{(2)} = ic_{2}, \quad c_{(3)} = ic_{3} , \quad c_{(4)} = -i.
\end{equation}
The number of linear independent solutions  given by (\ref{6.14}), (\ref{6.15}) and (\ref{6.16}) reads
\begin{eqnarray}\label{6.17}
N = 32\mathcal{N}= n_{1}(-ic_{1})n_{2}(-ic_{2})n_{3}(-ic_{3})n_{4}(i) +  n_{1}(i c_{1})n_{2}(ic_{2})n_{3}(ic_{3})n_{4}(-i),
\end{eqnarray}
where $n_{j}(c_{j})$ is the number of chiral parallel spinors on $M_{j}$, $j=1,2,3,4$.

{\bf Examples.}

 Let $M_{0} = \mathbb{R}$ and $M_1 = M_2 = M_3 = \mathbb{R}^2$.  Then all $n_j (c) = 1$,
 $j = 1,2,3$, with $c = \pm i$, and hence we get from (\ref{6.17})
   \begin{equation}\label{6.18}
  N =  32{\cal N}  = n_{4}(i) + n_{4}(-i).
  \end{equation}
\begin{description}
  \item[\{a\}] If we take the common worldvolume to be Minkowski space $M_{4} = \mathbb{R}^{1,3}$, then we have
$n_{4}(c) = 2$ and hence $\mathcal{N} = 1/8$ for all values of $c_i$, $i = 1,2,3$.
  \item[\{b\}] Now consider the case of $M5$-branes intersecting by the manifold  $M_{4} = (\mathbb{R}^{1,1}_{*}/Z_{2}) \times \mathbb{R}^{2}$,
  where $\mathbb{R}^{1,1}_{*} =  \mathbb{R}^{1,1}   \setminus  \{ {\bf 0 } \}$. Since $\mathbb{R}^{1,1}_{*}$ has only one parallel spinor
(left or right) depending of the choice of the spin structure, one obtains $n_{4}(c) = 1$ for any $c$ and hence $\mathcal{N} = 1/16$.
  \item[\{c\}] The same result takes place when $(M_{4},g^{4})$ is a 4-dimensional Ricci-flat $pp$-wave space from \cite{FOF} with the holonomy group $H = \mathbb{R}^{2}$ (see \cite{Bohle}). In this case $(n_{4}(i), n_{4}(-i)) = (1,1)$ and $\mathcal{N} = 1/16$.
\end{description}

 \textbf{(ii)}

The second possible configuration of three $M5$-branes is the pairwise intersection over 3-branes defined on the manifold of the form
  \begin{eqnarray}\label{7.1}
M_{0}\times M_{1}\times M_{2}\times M_{3}\times M_{4},
\end{eqnarray}
where $d_{0} = 3$, $d_{1}=d_{2}=d_{3}=d_{4}=2$. In Figure 3 one
presents the intersection of three magnetic branes.

\begin{figure}[!h]
\begin{eqnarray}
g = \left\{
\begin{array}{ccccccccccccc}
- & - & - & - &  - & \times & \times & \times & \times & \times &\times & &:H_1 \nonumber \\
- & - & - & \times & \times & - & - & \times & \times & \times &\times & &:H_2  \\
- & - &\hbox to 0pt{\hss$\underbrace{\hskip1.6cm}_{M_0}$\hskip 1cm\hss} - & \times &\hbox to 0pt{\hss$\underbrace{\hskip 1cm}_{M_1}$\hskip 4mm\hss}\times&\times &\hbox to 0pt{\hss$\underbrace{\hskip 1cm}_{M_2}$\hskip 4mm\hss} \times& - & \hbox to 0pt{\hss$\underbrace{\hskip 1cm}_{M_3}$\hskip 4mm\hss}- & \times & \hbox to 0pt{\hss$\underbrace{\hskip 1cm}_{M_4}$\hskip 4mm\hss}\times  & &:H_3.
\end{array} \right.
\end{eqnarray}
 \caption{The pairwise intersection of three $M5$-branes over 3-branes. $M_{0}$ is the totally transverse space, $M_{1}$,$M_{2}$,$M_{3}$ are the relative transverse spaces and $M_{4}$ is the common worldvolume.}
\end{figure}

The metric and the 4-form field strength  can be represented in the following form
\begin{eqnarray}\label{7.2}
g = H^{2/3}_{1}H^{2/3}_{2}H^{2/3}_{3}\Bigl\{g^{0} + H^{-1}_{2}H^{-1}_{3}g^{1} + H^{-1}_{1}H^{-1}_{3}g^{2} + H^{-1}_{1}H^{-1}_{2}g^{3} + H^{-1}_{1}H^{-1}_{2}H^{-1}_{3}g^{4}\Bigr\}, \\ \label{7.2A}
F = c_{1}(\ast_{0}dH_{1})\wedge\tau_{1} + c_{2}(\ast_{0}dH_{2})\wedge \tau_{2} + c_{3}(\ast_{0} dH_{3})\wedge \tau_{3},
\end{eqnarray}
where $c^{2}_{1} = c^{2}_{2} = c^{2}_{3} = 1$; $H_{1}$, $H_{2}$,
$H_{3}$ are harmonic functions on $(M_{0},g^{0})$. The metrics
$g^{i}$, $i=0,1,2,3$, have Euclidean signatures and the metric
$g^{4}$ has the signature $(-,+)$. The branes sets are $I_{1} = \{2,3,4\}$, $I_{2} = \{1,3,4 \}$, $I_{3} = \{1,2,4\}$.

The gamma matrices can be split in the following form
\begin{eqnarray}\label{7.3}
(\hat{\Gamma}^{A})=
\begin{array}{ccccccccc}
  (i\hat{\Gamma}^{a_{0}}_{(0)} & \otimes & \hat{\Gamma}_{(1)} & \otimes & \hat{\Gamma}_{(2)} & \otimes & \hat{\Gamma}_{(3)} & \otimes & \hat{\Gamma}_{(4)}, \\
  \textbf{1}_{2} & \otimes &  \hat{\Gamma}^{a_{1}}_{(1)} & \otimes &  \hat{\Gamma}_{(2)} & \otimes &  \hat{\Gamma}_{(3)} & \otimes &  \hat{\Gamma}_{(4)}, \\
  \textbf{1}_{2} & \otimes &  \mathbf{1}_{2} & \otimes &  i\hat{\Gamma}^{a_{2}}_{(2)} & \otimes & \hat{\Gamma}_{(3)} & \otimes &  \hat{\Gamma}_{(4)}, \\
  \textbf{1}_{2} & \otimes &  \mathbf{1}_{2} & \otimes &  \mathbf{1}_{2} & \otimes & \hat{\Gamma}^{a_{3}}_{(3)} & \otimes &  \hat{\Gamma}_{(4)}, \\
 \textbf{1}_{2} & \otimes &  \mathbf{1}_{2} & \otimes & \mathbf{1}_{2} & \otimes & \mathbf{1}_{2} & \otimes &  \hat{\Gamma}^{a_{4}}_{(4)}), \\
\end{array}
\end{eqnarray}
where
\begin{eqnarray}\label{7.4}
\hat{\Gamma}_{(0)} = \hat{\Gamma}^{1_{0}}_{(0)}\hat{\Gamma}^{2_{0}}_{(0)}\hat{\Gamma}^{3_{0}}_{(0)}, \quad \hat{\Gamma}_{(1)} = \hat{\Gamma}^{1_{1}}_{(1)}\hat{\Gamma}^{2_{1}}_{(1)}
\quad \hat{\Gamma}_{(2)} = \hat{\Gamma}^{1_{2}}_{(2)}\hat{\Gamma}^{2_{2}}_{(2)}, \nonumber\\
\hat{\Gamma}_{(3)} = \hat{\Gamma}^{1_{3}}_{(3)}\hat{\Gamma}^{2_{3}}_{(3)},
\quad \hat{\Gamma}_{(4)} = \hat{\Gamma}^{1_{4}}_{(4)}\hat{\Gamma}^{2_{4}}_{(4)}.
\end{eqnarray}
satisfy
\begin{equation}\label{7.4a}
(\hat{\Gamma}_{(0)})^{2} = (\hat{\Gamma}_{(1)})^{2}
=(\hat{\Gamma}_{(2)})^{2} = (\hat{\Gamma}_{(3)})^{2} =
-\textbf{1}_{2},\quad (\hat{\Gamma}_{(4)})^{2} = \textbf{1}_{2}.
\end{equation}

Here $\hat{\Gamma}^{a_{i}}_{(i)}$ are $2 \times 2$ gamma matrices
corresponding to $M_{i}$, $i = 0,1, 2, 3, 4. $  One can write down the gamma matrices corresponding to $M_{0}$ as
$\left(\hat{\Gamma}^{a_{0}}_{(0)}\right) = (\sigma_{1},
\sigma_{2}, \sigma_{3})$ and hence
\begin{equation}\label{7.4b}
\hat{\Gamma}_{(0)} = i\textbf{1}_{2}.
\end{equation}

Here we put
\begin{equation}\label{7.8}
\eta = \eta_{0}(x)\otimes \eta_{1}(y_{1})\otimes\eta_{2}(y_{2})\otimes \eta_{3}(y_{3})\otimes\eta_{4}(y_{4}),
\end{equation}
where $\eta_{0} = \eta_{0}(x)$ is a 2-component spinor on $M_{0}$, $\eta_{i} = \eta_{i}(y_{i})$ is a 2-component spinor on $M_{i}$,$i=1,2,3,4$.

The factorization relations (\ref{5.7a}) are valid for spinorial covariant derivatives where
\begin{eqnarray}\label{7.10}
\bar{D}_{m_{0}} = \partial_{m_{0}} + \frac 14 \omega^{(0)}_{a_{0}b_{0}m_{0}}\left(\hat{\Gamma}^{a_{0}}_{(0)}\hat{\Gamma}^{b_{0}}_{(0)}\otimes \textbf{1}_{2} \otimes \textbf{1}_{2}\otimes \textbf{1}_{2}\otimes \textbf{1}_{2}\right),\nonumber\\
...      \nonumber\\
\bar{D}_{m_{4}} = \partial_{m_{4}} + \frac 14
\omega^{(4)}_{a_{4}b_{4}m_{4}}\left(\textbf{1}_{2}\otimes
\textbf{1}_{2}\otimes \textbf{1}_{2}\otimes \textbf{1}_{2} \otimes
\hat{\Gamma}^{a_{4}}_{(4)}\hat{\Gamma}^{b_{4}}_{(4)}\right),
\end{eqnarray}
$D^{(i)}_{m_{i}}$ is a covariant derivative corresponding to $M_{i}$, $i=0,1,2,3,4$.

Due to (\ref{7.3}) and  (\ref{7.4})-(\ref{7.4b})
the brane operators corresponding to the magnetic branes can be written down as
\begin{equation}\label{7.5}
\hat{\Gamma}_{[s]}= \hat{\Gamma}^{1_{0}}\hat{\Gamma}^{2_{0}}\hat{\Gamma}^{3_{0}}\hat{\Gamma}^{1_{1}}\hat{\Gamma}^{2_{1}} =
\textbf{1}_{2} \otimes\textbf{1}_{(2)} \otimes \hat{\Gamma}_{(2)}\otimes \hat{\Gamma}_{(3)}\otimes\hat{\Gamma}_{(4)}
\end{equation}
for $s= I_{1}$,
\begin{equation}\label{7.6}
\hat{\Gamma}_{[s]} = \hat{\Gamma}^{1_{0}}\hat{\Gamma}^{2_{0}}\hat{\Gamma}^{3_{0}}\hat{\Gamma}^{1_{2}}\hat{\Gamma}^{2_{2}} =
\textbf{1}_{2} \otimes\hat{\Gamma}_{(1)}\otimes \textbf{1}_{2} \otimes \hat{\Gamma}_{(3)}\otimes \hat{\Gamma}_{(4)}
\end{equation}
 for  $s = I_{2}$,
\begin{equation}\label{7.7}
\hat{\Gamma}_{[s]} =\hat{\Gamma}^{1_{0}}\hat{\Gamma}^{2_{0}}\hat{\Gamma}^{3_{0}}\hat{\Gamma}^{1_{3}}\hat{\Gamma}^{2_{3}} =
\textbf{1}_{2} \otimes\hat{\Gamma}_{(1)}\otimes \hat{\Gamma}_{(2)} \otimes \textbf{1}_{2} \otimes\hat{\Gamma}_{(4)}
\end{equation}
for $s = I_{3}$.

The supersymmetry restrictions (\ref{2.14}) are satisfied if
\begin{eqnarray}\label{7.11}
\hat{\Gamma}_{(1)}\eta_{1} = c_{1} \eta_{1}, \qquad c^{2}_{(1)} = 1,\\
\hat{\Gamma}_{(j)}\eta_{j} = c_{(j)} \eta_{j}, \qquad c^{2}_{(j)} = -1,
\end{eqnarray}
with $j=0,2,3,4$.
Then  the conditions for the chirality constants are given by
\begin{equation}\label{7.12}
c_{(1)}c_{(3)}c_{(4)} = c_{1}, \quad c_{(1)}c_{(2)}c_{(3)} = c_{2}, \quad c_{(1)}c_{(2)}c_{(4)} = c_{3}.
\end{equation}
We get the following solution to Eqs. (\ref{2.6}) corresponding to the field configuration from (\ref{7.2})-(\ref{7.2A})
\begin{equation}\label{7.13}
\varepsilon = H^{-1/12}_{1} H^{-1/12}_{2} H^{-1/12}_{3} \eta_{0}(x) \otimes \eta_{1}(y_{1}) \otimes \eta_{2}(y_{2})\otimes\eta_{2}(y_{3})\otimes\eta_{4}(y_{4}).
\end{equation}
Here $\eta_{i}$, $i=0,1,2,3,4$, are chiral parallel spinors defined on $M_{i}$ $\left(D^{(i)}_{m_{i}}\eta_{i} = 0\right)$, obeying (\ref{7.11})-(\ref{7.12}).
Eqs. (\ref{7.12}) have the following solutions
\begin{equation}\label{7.14}
c_{(1)} = \pm ic_{2}c_{3}, \quad c_{(2)} = \pm ic_{1}c_{2}, \quad c_{(3)} = \pm i , \quad c_{(4)} =   -c_{1}c_{2}c_{3}.
\end{equation}
The number of linear independent solutions  is
\begin{equation}\label{7.15}
N = 32\mathcal{N}= n_{0}n_{4}(-c_{1}c_{2}c_{3})\sum_{c=\pm1}n_{1}(i c c_{2}c_{3})n_{2}(ic c_{1}c_{2})n_{3}(ic),
\end{equation}
where $n_{j}(c_{j})$ is the number of chiral parallel spinors on $M_{j}$, $j=1,2,3,4$, $n_{0}$ is the number of parallel spinors on $M_{0}$.

\textbf{Examples.}

 Consider the case when all factor spaces are
flat:  $M_{0} = \mathbb{R}^{3}$, $M_{1} = M_{2} = M_{3} =
\mathbb{R}^{2}$. Then all $n_{j}(ic) = 1$, $j=1,2,3$, $c = \pm 1$.
The amount of preserved supersymmetries is given by
\begin{equation}\label{7.16}
\mathcal{N} = \frac{1}{8}n_{4}(-c_{1}c_{2}c_{3}).
\end{equation}
\begin{description}
  \item[\{a\}] Let $M_{4} = \mathbb{R}^{1,1}$, then we obtain $\mathcal{N} = \displaystyle{\frac{1}{8}}$ for any choice of brane sign factors.
  \item[\{b\}] If $M_{4} = \mathbb{R}^{1,1}_{*}/Z_{2}$ with $n_4(1) =1$ and $n_4(-1) =0$, we get $\mathcal{N} = \displaystyle{\frac{1}{8}}$ for $c_{1}c_{2}c_{3} = -1$ and $\mathcal{N} = 0$ otherwise.
\end{description}

\textbf{(iii)}

 The third intersection of magnetic branes is defined on the manifold
  \begin{eqnarray}\label{8.1}
M_{0}\times M_{1}\times M_{2}\times M_{3}\times M_{4}\times M_{5}\times M_{6}\times M_{7},
\end{eqnarray}
where $d_{0} = 2$, $d_{1}= d_{2} = d_{3}=d_{4}= d_{5} = d_{6}=1$, $d_{7} =3$. For this configuration we have Fig.4.

\begin{figure}[!h]
\begin{eqnarray}
g = \left\{
\begin{array}{ccccccccccccc}
- & - & \times & - &  - & - & \times & \times & \times & \times &\times & &:H_1 \nonumber \\
- & - & - & \times & - & \times & - & \times & \times & \times &\times & &:H_2  \\
- &  \hbox to 0pt{\hss$\underbrace{\hskip 1cm}_{M_0}$\hskip 4mm\hss}- & \underbrace{-}_{M_{1}} &\underbrace{-}_{M_{2}} & \underbrace{\times}_{M_{3}} &\underbrace{\times}_{M_{4}}& \underbrace{\times}_{M_{5}} & \underbrace{-}_{M_{6}}& \times & \times & \hbox to 0pt{\hss$\underbrace{\hskip1.6cm}_{M_7}$\hskip 1.1cm\hss}\times  & &:H_3.
\end{array} \right.
\end{eqnarray}
 \caption{The pairwise intersection of three $M5$-branes over 3-branes. $M_{0}$ is the totally transverse space, $M_{i}$,$i=1,2,3,4,5,6,$  are the relative transverse spaces and $M_{7}$ is the common worldvolume.}
\end{figure}

The metric of three intersecting $M5$-branes now reads
\begin{eqnarray}\label{8.2}
g = H^{2/3}_{1}H^{2/3}_{2}H^{2/3}_{3}\Bigl\{g^{0} +  H^{-1}_{1} g^{1} +  H^{-1}_{2}g^{2} + H^{-1}_{3}g^{3} + H^{-1}_{2}H^{-1}_{3}g^{4} + H^{-1}_{1}H^{-1}_{3}g^{5}+ \nonumber\\
H^{-1}_{1} H^{-1}_{2}g^{6} +  H^{-1}_{1}H^{-1}_{2}H^{-1}_{3}g^{7} \Bigr\}.
\end{eqnarray}
The corresponding field strength is
\begin{equation}\label{8.2A}
F = c_{1}(\ast_{0}dH_{1})\wedge\tau_{2}\wedge\tau_{3}\wedge\tau_{4} + c_{2}(\ast_{0}dH_{2})\wedge\tau_{1}\wedge\tau_{3}\wedge\tau_{5} +
 c_{3}(\ast_{0}dH_{3})\wedge \tau_{1}\wedge \tau_{2}\wedge \tau_{6},
\end{equation}
where $c^{2}_{1} = c^{2}_{2} = c^{2}_{3} = 1$; $H_{1}$, $H_{2}$, $H_{3}$ are harmonic functions on $(M_{0},g^{0})$. The metrics $g^{i}$, $i=0,1,2,3,4,5,6$, have Euclidean signatures and the metric $g^{7}$ has the signature $(-,+,+)$. The branes sets are $I_{1} = \{1,5,6,7\}$, $I_{2} = \{2,4,6,7\}$, $I_{3} = \{3,4,5,7\}$.

Under the decomposition rules the set of gamma matrices can be presented in the form
\begin{eqnarray}\label{8.3}(\hat{\Gamma}^{A})=
  \begin{array}{ccccccccccccccccccccc}
(\hat{\Gamma}^{a_{0}}_{(0)} &\otimes & 1 & \otimes & 1 & \otimes & 1 & \otimes & 1 & \otimes & 1 & \otimes & 1 & \otimes & \textbf{1}_{2} & \otimes & \textbf{1}_{2}& \otimes & \textbf{1}_{2}& \otimes & \textbf{1}_{2},\\
i\hat{\Gamma}_{(0)} & \otimes & 1 & \otimes & 1 & \otimes & 1 & \otimes & 1 & \otimes & 1 & \otimes& 1 & \otimes & \hat{\Gamma}_{(7)} & \otimes & \sigma_{3}  & \otimes & \textbf{1}_{2} & \otimes & \textbf{1}_{2},\\
i\hat{\Gamma}_{(0)} & \otimes & 1 & \otimes & 1 & \otimes & 1 & \otimes & 1 & \otimes & 1 & \otimes & 1 & \otimes & \hat{\Gamma}_{(7)} & \otimes & \sigma_{1} & \otimes & \textbf{1}_{2} & \otimes & \textbf{1}_{2}, \\
i\hat{\Gamma}_{(0)} & \otimes & 1 & \otimes & 1 & \otimes & 1 & \otimes & 1 & \otimes & 1 & \otimes & 1 & \otimes & \hat{\Gamma}_{(7)} & \otimes & \sigma_{2}  & \otimes & \sigma_{3}  & \otimes & \textbf{1}_{2}, \\
i\hat{\Gamma}_{(0)} & \otimes & 1 & \otimes & 1 & \otimes & 1 & \otimes & 1 & \otimes & 1 & \otimes & 1 & \otimes & \hat{\Gamma}_{(7)} & \otimes & \sigma_{2} & \otimes & \sigma_{1} & \otimes & \textbf{1}_{2}, \\
i\hat{\Gamma}_{(0)} & \otimes & 1 & \otimes & 1 & \otimes & 1 & \otimes & 1 & \otimes &  1 & \otimes & 1 & \otimes & \hat{\Gamma}_{(7)} & \otimes & \sigma_{2} & \otimes & \sigma_{2} & \otimes &  \sigma_{3}, \\
i\hat{\Gamma}_{(0)} & \otimes & 1 & \otimes & 1 & \otimes & 1 & \otimes & 1 & \otimes & 1 & \otimes & 1 & \otimes & \hat{\Gamma}_{(7)} & \otimes & \sigma_{2} & \otimes & \sigma_{2} & \otimes & \sigma_{1}, \\
i\hat{\Gamma}_{(0)} & \otimes & 1 & \otimes & 1  & \otimes & 1 & \otimes & 1 & \otimes & 1 & \otimes & 1 & \otimes & \hat{\Gamma}^{a_{7}}_{(7)} & \otimes & \sigma_{2} & \otimes & \sigma_{2} & \otimes & \sigma_{2}).
  \end{array}
\end{eqnarray}
Here the operators
\begin{equation}\label{8.4}
\hat{\Gamma}_{(0)} = \hat{\Gamma}^{1_{0}}_{(0)}\hat{\Gamma}^{2_{0}}_{(0)}, \quad \hat{\Gamma}_{(7)} = \hat{\Gamma}^{1_{7}}_{(7)}\hat{\Gamma}^{2_{7}}_{(7)}\hat{\Gamma}^{3_{7}}_{(7)}
\end{equation}
obey
\begin{equation}\label{8.4A}
(\hat{\Gamma}_{(0)})^{2} =- \textbf{1}_{2}, \quad (\hat{\Gamma}_{(7)})^{2} = \textbf{1}_{2}.
\end{equation}
The gamma matrices corresponding to $M_{0}$ and $M_{7}$ manifolds can be written down in the form
$\left(\hat{\Gamma}^{a_{0}}_{(0)}\right) = \left(\sigma_{1},\sigma_{2}\right)$, $\hat{\Gamma}_{(0)} = i\sigma_{3}$, $\left(\hat{\Gamma}^{a_{7}}_{(7)}\right) = \left(i\sigma_{1},\sigma_{2}, \sigma_{3}\right)$, $\hat{\Gamma}_{(7)} = - \textbf{1}_{2}$, respectively.

We put the following relation for the 32-component spinor
 \begin{equation}\label{8.4b}
 \eta = \eta_{0}(x) \otimes \eta_{1}(y_{1}) \otimes
\eta_{2}(y_{2}) \otimes \eta_{3}(y_{3})  \otimes \eta_{4}(y_{4})
\otimes \eta_{5}(y_{5}) \otimes \eta_{6}(y_{6}) \otimes
\eta_{7}(y_{7}) \otimes \chi,
\end{equation}
where $\eta_{i} = \eta_{i}(y_{i})$ is a 1-component spinor on
$M_{i}$,$i=1,2,3,4,5,6$, $\eta_{0} = \eta_{0}(x)$ is a
2-component spinor on $M_{0}$,  $\eta_{7} = \eta_{7}(y_{7})$ is a
2-component spinor on $M_{7}$ and
 \begin{equation}\label{8.4c}
 \chi = \sum_{r =1}^{8} \chi^r_{1} \otimes \chi^r_{2} \otimes \chi^r_{3}
 \end{equation}
  belongs to $V = \mathbb{C}^{2}\otimes \mathbb{C}^{2}\otimes
  \mathbb{C}^{2}$; $\chi^r_{1},  \chi^r_{2},  \chi^r_{3}$ are ``auxiliary''
  2-dimensional spinors.

We remind that auxiliary spinors appear when the dimension of the spinorial space  $2^{[D/2]}$
corresponding to the product manifold $M$ is not equal to the product of dimensions  of spinorial spaces $2^{[d_l/2]}$ corresponding to factor spaces $M_l$
(or, equivalently, when the integer part $[D/2]$ is not equal to the sum of integer parts $[d_l/2]$).
The simplest example of the product manifold with two factor spaces of odd dimensions $d_1$ and $d_2$ was considered in \cite{LPR}.
In this case the auxiliary spinor is two-dimensional one.

Here the covariant derivatives act on  $\eta$ as
\begin{eqnarray}\label{8.9}
\bar{D}_{m_{0}}\eta = \left(D^{(0)}_{m_{0}}\eta_{0}\right)\otimes\eta_{1}\otimes\eta_{2}\otimes\eta_{3}\otimes \eta_{4}\otimes \eta_{5}\otimes\eta_{6}\otimes\eta_{7}\otimes\chi,\nonumber \\
  ...   \nonumber \\
\bar{D}_{m_{7}}\eta = \eta_{0}\otimes\eta_{1}\otimes\eta_{2}\otimes\eta_{3}\otimes \eta_{4}\otimes\eta_{5}\otimes \eta_{6}\otimes \left(D^{(7)}_{m_{7}}\eta_{7}\right)\otimes\chi,
\end{eqnarray}
where
\begin{eqnarray}\label{8.10}
\bar{D}_{m_{0}} = \partial_{m_{0}} + \frac 14 w^{(0)}_{a_{0}b_{0}m_{0}}\left(\hat{\Gamma}^{a_{0}}_{(0)}\hat{\Gamma}^{b_{0}}_{(0)}\otimes 1\otimes 1 \otimes 1\otimes 1\otimes 1 \otimes 1\otimes \mathbf{1}_{2} \otimes \mathbf{1}_{2}  \otimes  \mathbf{1}_{2} \otimes \mathbf{1}_{2}\right),\nonumber \\
\bar{D}_{m_{7}} = \partial_{m_{7}} + \frac 14 w^{(7)}_{a_{7}b_{7}m_{7}}\left(\mathbf{1}_{2}\otimes 1 \otimes 1 \otimes 1\otimes 1  \otimes 1 \otimes 1\otimes \hat{\Gamma}^{a_{7}}_{(7)}\hat{\Gamma}^{b_{7}}_{(7)}\otimes \mathbf{1}_{2}\otimes \mathbf{1}_{2} \otimes \mathbf{1}_{2}\right),
\end{eqnarray}
$\bar{D}_{m_{i}} = D^{(i)}_{m_{i}} = \partial_{m_{i}}$, $i=1,2,3,4,5,6$.
$D^{(i)}_{m_{i}}$ is a covariant derivative corresponding to $M_{i}$, $i=0,7$. Using (\ref{8.3}) one can write down the
operators for $M5$-branes as
\begin{equation}\label{8.7a}
 \hat{\Gamma}_{[s]}= \hat{\Gamma}^{1_{0}}\hat{\Gamma}^{2_{0}}\hat{\Gamma}^{1_{2}}\hat{\Gamma}^{1_{3}}\hat{\Gamma}^{1_{4}} =
   - \textbf{1}_{2} \otimes 1 \otimes 1 \otimes 1 \otimes 1 \otimes 1 \otimes 1 \otimes \hat{\Gamma}_{(7)} \otimes B_1
 \end{equation}
 for $s = I_{1}$,
 \begin{equation}\label{8.7b}
 \hat{\Gamma}_{[s]} =
 \hat{\Gamma}^{1_{0}}\hat{\Gamma}^{2_{0}}\hat{\Gamma}^{1_{1}}\hat{\Gamma}^{1_{3}}\hat{\Gamma}^{1_{5}}
  = - \textbf{1}_{2} \otimes 1 \otimes 1 \otimes 1 \otimes 1 \otimes 1 \otimes 1 \otimes \hat{\Gamma}_{(7)} \otimes B_2
\end{equation}
for $s = I_{2}$,
\begin{equation}\label{8.7c}
\hat{\Gamma}_{[s]}
=\hat{\Gamma}^{1_{0}}\hat{\Gamma}^{2_{0}}\hat{\Gamma}^{1_{1}}\hat{\Gamma}^{1_{2}}\hat{\Gamma}^{1_{6}}
= - \textbf{1}_{2} \otimes 1 \otimes 1 \otimes 1 \otimes 1 \otimes
1 \otimes 1 \otimes \hat{\Gamma}_{(7)} \otimes B_3
\end{equation}
for $s = I_{3}$.

Here we denote by $B_s$ the following self-adjoint commuting
idempotent  (i.e. $B_s^2 =   \textbf{1}_{2}\otimes \textbf{1}_{2}
\otimes \textbf{1}_{2}$) operators acting on the 8-dimensional
Hilbert space $V= \mathbb{C}^{2}\otimes \mathbb{C}^{2}\otimes\mathbb{C}^{2}$
 \begin{equation}\label{8.7d}
 B_1 = - \sigma_{1} \otimes \sigma_{2}  \otimes \textbf{1}_{2},
  \quad
 B_2 =  \sigma_{3} \otimes \sigma_{1}  \otimes  \sigma_{3},
   \quad
 B_3 = - \textbf{1}_{2}  \otimes \sigma_{2}  \otimes
    \sigma_{1}.
 \end{equation}

 Due to the proposition from Appendix A there
  exists a basis of eigenvectors
 $ \psi_{\varepsilon_1,\varepsilon_2, \varepsilon_3} $ in $V$
   with $\varepsilon_1 = \pm 1, \varepsilon_2 = \pm 1, \varepsilon_3 = \pm 1$,
  obeying
  \begin{equation}\label{8.7e}
  B_s \psi_{\varepsilon_1, \varepsilon_2, \varepsilon_3} = \varepsilon_s \psi_{\varepsilon_1, \varepsilon_2,
  \varepsilon_3},
  \end{equation}
  $s = 1,2,3$.

 Let the gamma matrices corresponding to the 3-dimensional   manifold
 $M_{7}$ be chosen as follows $\hat{\Gamma}^{a_{7}}_{(7)} =
\left(i\sigma_{1},\sigma_{2}, \sigma_{3}\right)$, and hence
$\hat{\Gamma}_{(7)} = - \textbf{1}_{2}$.

 Then the solutions to SUSY equations (\ref{2.6})
 corresponding to the field configuration from (\ref{8.2}), (\ref{8.2A})
  are generated by the following set of monomial solutions
  \begin{equation}\label{8.7f}
  \varepsilon  = H_1^{-1/12} H_2^{-1/12} H_3^{-1/12}
  \eta_0(x) \otimes \eta_1 \otimes \eta_2 \otimes \eta_3 \otimes
  \eta_4 \otimes \eta_5 \otimes \eta_6 \otimes \eta_7(y_7) \otimes
  \psi_{\varepsilon_1, \varepsilon_2, \varepsilon_3},
 \end{equation}
 where  $\eta_0(x)$ and  $\eta_7(y_7)$ are  parallel spinors
 defined on $M_0$ and $M_7$, respectively,
  $\eta_i$ are constant  1-dimensional spinors, $i
  =1,2,3,4,5,6$. Here $\varepsilon_s$ parameters obey
  the relations
   \begin{equation}\label{8.8b}
      \varepsilon_s = c_s,
     \end{equation}
 $s = 1, 2, 3$,
  following from relations (\ref{2.14}), (\ref{8.4b}) with $\chi = \psi_{\varepsilon_1, \varepsilon_2, \varepsilon_3}$,
  (\ref{8.7a}),   (\ref{8.7b}), (\ref{8.7c}) and (\ref{8.7e}).

 The number of linear independent solutions
  given by (\ref{8.7f}) and (\ref{8.8b}) is

  \begin{equation}\label{8.8c}
   N  = 32 {\cal N} =  n_0 n_7,
   \end{equation}
 where $n_j$ is the number of  parallel spinors on
 $M_j$, $j = 0,7$.

\textbf{Examples.} \begin{description}
                     \item[\{a\}] For $M_{0} = \mathbb{R}^{2}$ and $M_{7} = \mathbb{R}^{1,2}$ we get from (\ref{8.8c}) ${\cal N} = 1/8$ in agreement with \cite{BREJS}.
                     \item[\{b\}] If we put $M_{7} = (\mathbb{R}^{1,1}_{*}/Z_{2})\times \mathbb{R}$ instead of the 3-dimensional analogue of  Minkowski space $M_{7} = \mathbb{R}^{1,2}$ one obtains ${\cal N} = 1/16$. This result does not depend upon the brane sign factors.
                   \end{description}

\subsection{$M2\cap M2 \cap M5$}

Let us consider the composite configuration of two electric branes
each intersecting $M5$-brane over a string with  the  two  strings
intersecting  over  a point.
 $M2\cap M2\cap M5$-solution is defined on the manifold
\begin{eqnarray}\label{9.1}
M_{0}\times M_{1}\times M_{2}\times M_{3}\times M_{4}\times M_{5}\times M_{6},
\end{eqnarray}
where $d_{0} = 3$, $d_{1}=d_{2} = d_{4}=d_{5} = d_{6}= 1$ and $d_{3}= 3$.

The intersection is given in Fig. 5.

\begin{figure}[!h]
\begin{eqnarray}
g = \left\{
\begin{array}{ccccccccccccc}
- & - & - & \times & - & - & - & - & - & \times &\times & &:H_1 \nonumber \\
- & - & - & - & \times & - & - & - & \times & - &\times & &:H_2  \\
- & - & \hbox to 0pt{\hss$\underbrace{\hskip1.6cm}_{M_0}$\hskip 1cm\hss} -&\underbrace{-}_{M_{1}} & \underbrace{-}_{M_{2}} & \times & \times & \hbox to 0pt{\hss$\underbrace{\hskip1.6cm}_{M_3}$\hskip 1cm\hss}\times & \underbrace{\times}_{M_4} & \underbrace{\times}_{M_5} & \underbrace{\times}_{M_6}  & &:H_3.
\end{array} \right.
\end{eqnarray}
 \caption{$M2\cap M2\cap M5$: $M2$-branes intersect over a point, each $M2$-brane intersects  $M5$-brane over a string, $M_{0}$ is the totally transverse space,$M_{i}$, $i=1,2,3,4,5$, are the relative transverse spaces, $M_{6}$ is the common worldvolume.}
\end{figure}

The metric and the 4-form field strength corresponding to the intersection of two $M2$-branes
and one $M5$-brane can be represented in the following form
\begin{eqnarray}\label{9.2}
g = H^{1/3}_{1}H^{1/3}_{2}H^{2/3}_{3}\Bigl\{g^{0} + H^{-1}_{1}g^{1} +  H^{-1}_{2}g^{2} + H^{-1}_{3}g^{3}
 + H^{-1}_{2}H^{-1}_{3}g^{4} + \nonumber \\
H^{-1}_{1}H^{-1}_{3}g^{5} + H^{-1}_{1}H^{-1}_{2}H^{-1}_{3}g^{6} \Bigr\}, \\
\label{9.2A}
F = c_{1}dH^{-1}_{1}\wedge \tau_{1}\wedge \tau_{5}\wedge \tau_{6} +
c_{2}dH^{-1}_{2}\wedge \tau_{2}\wedge \tau_{4}\wedge\tau_{6} +
 c_{3}(\ast_{0} dH_{3})\wedge \tau_{1}\wedge \tau_{2},
\end{eqnarray}
where $c^{2}_{1} = c^{2}_{2} = c^{2}_{3} = 1$; $H_{1}$, $H_{2}$,
$H_{3}$ are harmonic functions on $(M_{0},g^{0})$. The metrics
$g^{i}$, $i=0,1,2,3,4,5$, have Euclidean signatures and we put the
metric $g^{6} = -dt\otimes dt$. The brane sets are $I_{1} =
\{1,5,6\}$, $I_{2} = \{2,4,6\}$, $I_{3} = \{3,4,5,6\}$.

The gamma matrices may be chosen  in the following form
\begin{eqnarray}\label{9.3}
(\hat{\Gamma}^{A})=
(\hat{\Gamma}^{a_{0}}_{(0)} \otimes 1\otimes 1 \otimes \textbf{1}_{2} \otimes 1  \otimes 1 \otimes 1 \otimes \sigma_{3}\otimes  \textbf{1}_{2}\otimes \textbf{1}_{2},\nonumber \\
\textbf{1}_{2} \otimes 1 \otimes 1 \otimes \textbf{1}_{2} \otimes 1  \otimes 1 \otimes 1 \otimes \sigma_{1} \otimes  \textbf{1}_{2}\otimes \textbf{1}_{2}, \nonumber\\
\textbf{1}_{2} \otimes 1 \otimes 1 \otimes \textbf{1}_{2} \otimes 1 \otimes 1 \otimes 1 \otimes \sigma_{2}  \otimes \sigma_{3}\otimes \textbf{1}_{2}, \nonumber \\
\textbf{1}_{2}  \otimes 1 \otimes 1  \otimes \hat{\Gamma}^{a_{3}}_{(3)} \otimes 1 \otimes 1 \otimes 1 \otimes \sigma_{2} \otimes \sigma_{1} \otimes \textbf{1}_{2}, \nonumber\\
\textbf{1}_{2}  \otimes 1 \otimes 1 \otimes \textbf{1}_{2}\otimes 1 \otimes 1 \otimes 1 \otimes \sigma_{2} \otimes \sigma_{2} \otimes \sigma_{3}, \nonumber \\
\textbf{1}_{2}  \otimes 1 \otimes 1  \otimes \textbf{1}_{2} \otimes 1 \otimes 1 \otimes 1 \otimes \sigma_{2} \otimes \sigma_{2}  \otimes \sigma_{1}, \nonumber\\
\textbf{1}_{2}  \otimes 1 \otimes 1  \otimes \textbf{1}_{2} \otimes 1  \otimes 1 \otimes i \otimes \sigma_{2} \otimes \sigma_{2} \otimes \sigma_{2}).
\end{eqnarray}
Here the operators
\begin{equation}\label{9.4}
\hat{\Gamma}_{(0)} = \hat{\Gamma}^{1_{0}}_{(0)}\hat{\Gamma}^{2_{0}}_{(0)}\hat{\Gamma}^{3_{0}}_{(0)}, \quad \hat{\Gamma}_{(3)} = \hat{\Gamma}^{1_{3}}_{(3)}\hat{\Gamma}^{2_{3}}_{(3)}\hat{\Gamma}^{3_{3}}_{(3)}
\end{equation}
obey
\begin{equation}\label{9.5}
(\hat{\Gamma}_{(0)})^{2} = (\hat{\Gamma}_{(3)})^{2} = - \textbf{1}_{2}.
\end{equation}

The  spinor monomial reads
\begin{equation}\label{9.5A}
\eta = \eta_{0}(x)
\otimes \eta_{1} (y_{1}) \otimes \eta_{2} (y_{2}) \otimes \eta_{3}(y_{3}) \otimes \eta_{4} (y_{4})
\otimes \eta_{5} (y_{5}) \otimes \eta_{6}(y_{6}) \otimes \chi,
\end{equation}
where $\eta_{i} = \eta_{i}(y_{i})$ is a 1-component spinor on
$M_{i}$, $i=1,2,4,5,6$, $\eta_{0} = \eta_{0}(x)$ is a 2-component
spinor on $M_{0}$,  $\eta_{3} = \eta_{3}(y_{3})$ is a 2-component
spinor on $M_{3}$ and $\chi$
belongs to $V = \mathbb{C}^{2}\otimes\mathbb{C}^{2}\otimes
\mathbb{C}^{2}$.

The  covariant derivatives $\bar{D}_{m_{i}}$ act on $\eta$ as follows
\begin{eqnarray}\label{9.6a}
\bar{D}_{m_{0}}\eta = \left(D^{(0)}_{m_{0}}\eta_{0}\right)\otimes\eta_{1}\otimes\eta_{2}\otimes\eta_{3}\otimes \eta_{4}\otimes \eta_{5}\otimes\eta_{6}\otimes\chi, \nonumber\\
...\nonumber\\
\bar{D}_{m_{3}}\eta = \eta_{0}\otimes\eta_{1}\otimes\eta_{2}\otimes\left(D^{(3)}_{m_{3}}\eta_{3}\right)\otimes \eta_{4}\otimes \eta_{5}\otimes\eta_{6}\otimes\chi,\nonumber\\
... \nonumber \\
\bar{D}_{m_{6}}\eta = \eta_{0}\otimes\eta_{1}\otimes\eta_{2}\otimes\eta_{3}\otimes\eta_{4}\otimes\eta_{5}\otimes\left(D^{(6)}_{m_{6}}\eta_{6}\right)\otimes\chi,
\end{eqnarray}
where
\begin{eqnarray}\label{9.6}
\bar{D}_{m_{0}} = \partial_{m_{0}} + \frac{1}{4}\omega^{(0)}_{a_{0}b_{0}m_{0}}
(\hat{\Gamma}^{a_{0}}_{(0)}\hat{\Gamma}^{b_{0}}_{(0)} \otimes  1\otimes1   \otimes \textbf{1}_{2} \otimes1  \otimes 1 \otimes  1 \otimes \textbf{1}_{2} \otimes \textbf{1}_{2}\otimes \textbf{1}_{2}),\nonumber\\
\bar{D}_{m_{3}} = \partial_{m_{3}} + \frac 14
\omega^{(3)}_{a_{3}b_{3}m_{3}}
(\textbf{1}_{2}\otimes 1\otimes 1 \otimes \hat{\Gamma}^{a_{3}}_{(3)}\hat{\Gamma}^{b_{3}}_{(3)}
  \otimes 1 \otimes 1 \otimes 1 \otimes \textbf{1}_{2}\otimes \textbf{1}_{2} \otimes \textbf{1}_{2}),
\end{eqnarray}
$\bar{D}_{m_{i}} = D^{(i)}_{m_{i}} = \partial_{m_{i}}$ for $i = 1,2,4,5,6$ and $D^{(i)}_{m_{i}}$ is a covariant derivative corresponding to $M_{i}$, $i=0,3$.

Under  (\ref{9.3}) the operators  corresponding to the $M2$-branes  read
\begin{eqnarray}\label{9.7}
\hat{\Gamma}_{[s]}= \hat{\Gamma}^{1_{1}}\hat{\Gamma}^{1_{5}}\hat{\Gamma}^{1_{6}} =
-\textbf{1}_{2}\otimes 1 \otimes 1\otimes \textbf{1}_{2}\otimes 1 \otimes 1\otimes 1 \otimes B_{1}
\end{eqnarray}
for $s= I_{1}$ and
\begin{eqnarray}\label{9.8}
\hat{\Gamma}_{[s]} =\hat{\Gamma}^{1_{2}}\hat{\Gamma}^{1_{4}}\hat{\Gamma}^{1_{6}} =
 - \textbf{1}_{2}\otimes 1 \otimes 1 \otimes\textbf{1}_{2} \otimes 1 \otimes 1 \otimes 1 \otimes B_{2}
\end{eqnarray}
for $s = I_{2}$. The operator
 for the $M5$-brane can be written in the form
\begin{eqnarray}\label{9.9}
\hat{\Gamma}_{[s]} = \hat{\Gamma}^{1_{0}}\hat{\Gamma}^{2_{0}}\hat{\Gamma}^{3_{0}}\hat{\Gamma}^{1_{1}}\hat{\Gamma}^{1_{2}} =
i \hat{\Gamma}_{(0)}\otimes 1 \otimes 1 \otimes  \textbf{1}_{2} \otimes 1 \otimes 1 \otimes 1 \otimes B_{3}
\end{eqnarray}
for $s = I_{3}$.

In (\ref{9.7})-(\ref{9.9}) $B_{s}$ are self-adjoint commuting
idempotent operators acting on $V =
\mathbb{C}^{2}\otimes\mathbb{C}^{2}\otimes\mathbb{C}^{2}$
\begin{equation}\label{9.9A}
B_{1} = \sigma_{1} \otimes \textbf{1}_{2} \otimes \sigma_{3}, \quad B_{2} = \sigma_{2} \otimes \sigma_{3} \otimes \sigma_{1}, \quad B_{3} = \textbf{1}_{2}\otimes \sigma_{3}\otimes \textbf{1}_{2}.
\end{equation}

The gamma matrices corresponding to $M_{0}$ can be chosen in the form $(\hat{\Gamma}^{a_{0}}_{(0)}) = \left(\sigma_{1},\sigma_{2}, \sigma_{3}\right)$ and hence
$\hat{\Gamma}_{(0)} = i\textbf{1}_{2}$.

Under the proposition  from  Appendix A there exists a basis of eigenvectors
$\psi_{\varepsilon_{1}\varepsilon_{2}\varepsilon_{3}}$ in $V$ with
$\varepsilon_{1} = \pm1$, $\varepsilon_{2} = \pm 1$,
$\varepsilon_{3} = \pm 1$, satisfying (\ref{8.7e}).

The solutions to generalized Killing equations (\ref{2.6}) corresponding to the field configuration from (\ref{9.2}), (\ref{9.2A}) are given by the following monomial solutions
\begin{equation}\label{9.10}
\varepsilon  = H^{-1/6}_{1}H^{-1/6}_{2}H^{-1/12}_{3}\eta_{0}(x)\otimes\eta_{1}\otimes \eta_{2}\otimes \eta_{3}(y_{3})\otimes \eta_{4}\otimes \eta_{5}\otimes \eta_{6}\otimes \psi_{\varepsilon_{1},\varepsilon_{2},\varepsilon_{3}},
\end{equation}
where $\eta_{0}(x)$ and $\eta_{3}(y_{3})$ are parallel spinors defined on $M_{0}$ and $M_{3}$, respectively, $\eta_{i}$ is a constant 1-dimensional spinor on $M_{i}$, $i =1,2,4,5,6$.

Using the relations (\ref{2.14}), (\ref{9.5A}) with $\chi = \psi_{\varepsilon_{1}, \varepsilon_{2}, \varepsilon_{3}}$, (\ref{9.7})-(\ref{9.9})
 one can obtain the restrictions for the parameters $\varepsilon_{s}$
\begin{equation}\label{9.11}
- \varepsilon_{s} = c_{s}.
\end{equation}
Thus  the  number of linear independent solutions given by (\ref{9.10}) and (\ref{9.11})
\begin{equation}\label{9.19}
N = 32\mathcal{N} = n_{0}n_{3},
\end{equation}
where $n_{j}$ is the number of parallel spinors on the
3-dimensional manifolds $M_{j}$,  $j=0,3$.

\textbf{Example.}

Let the totally transverse space $M_{0}$ and relatively transverse
space $M_{6}$ be coinciding with the 3-dimensional Euclidean
space: $M_{0} = M_{6} = \mathbb{R}^{3}$. Then one obtains $\mathcal{N}
 = 1/8$ in agreement with \cite{BREJS}.

\subsection{$M2\cap M5 \cap M5$}
According to the classification from \cite{BREJS} there are two possible configurations
for the intersection of one $M2$-brane and two $M5$-branes.

\textbf{(i)}

The first configuration $M2\cap M5\cap M5$ is defined on the manifold
  \begin{eqnarray}\label{10.1}
M_{0}\times M_{1}\times M_{2}\times M_{3}\times M_{4}\times M_{5},
\end{eqnarray}
where $d_0 = d_{2}=d_{3}=d_{4}=d_{5}=2$ and $d_{1}=1$ and describes a $M2$-brane intersecting each
of the two $M5$-branes over a string with the $M5$-branes intersecting over a 3-brane (see Fig. 6).

\begin{figure}[!h]
\begin{eqnarray}
g = \left\{
\begin{array}{ccccccccccccc}
- & - & \times & - &  - & - & - & - & - & \times &\times & &:H_1 \nonumber \\
- & - & - & \times & \times & - & - & \times & \times & \times &\times & &:H_2  \\
- & \hbox to 0pt{\hss$\underbrace{\hskip 1cm}_{M_0}$\hskip 4mm\hss}- & \underbrace{-}_{M_{1}} & -  &\hbox to 0pt{\hss$\underbrace{\hskip1cm}_{M_2}$\hskip 4mm\hss}- &\times &\hbox to 0pt{\hss$\underbrace{\hskip1cm}_{M_3}$\hskip 4mm\hss} \times& \times & \hbox to 0pt{\hss$\underbrace{\hskip 1cm}_{M_4}$\hskip 4mm\hss}\times & \times & \hbox to 0pt{\hss$\underbrace{\hskip 1cm}_{M_5}$\hskip 4mm\hss}\times  & &:H_3.
\end{array} \right.
\end{eqnarray}
 \caption{$M2\cap M5 \cap M5$-intersection. $M_{0}$ is the totally transverse space, $M_{1}$, $M_{2}$, $M_{3}$, $M_{4}$ are the relative transverse spaces and $M_{5}$ is the common worldvolume.}
\end{figure}

The solution for the intersection of an electric $M2$-brane and two magnetic $M5$-branes is given by
\begin{equation}\label{10.2}
g = H^{1/3}_{1}H^{2/3}_{2}H^{2/3}_{3}\Bigl\{g^{0} + H^{-1}_{1}g^{1} + H^{-1}_{2}g^{2} + H^{-1}_{3}g^{3}
+ H^{-1}_{2}H^{-1}_{3}g^{4} + H^{-1}_{1}H^{-1}_{2}H^{-1}_{3}g^{5} \Bigr\},
\end{equation}
\begin{equation}\label{10.3}
F = c_{1}dH^{-1}_{1}\wedge \tau_{1}\wedge \tau_{5} + c_{2}(\ast_{0}dH_{2})\wedge \tau_{1}\wedge \tau_{3} +
c_{3}(\ast_{0} dH_{3})\wedge \tau_{1}\wedge\tau_{2},
\end{equation}
where $c^{2}_{1} = c^{2}_{2} = c^{2}_{3} = 1$; $H_{1}$, $H_{2}$, $H_{3}$ are harmonic functions defined on $(M_{0},g^{0})$.
The metrics $g^{i}$, $i=0,1,2,3,4$, have Euclidean signatures and
the metric $g^{5}$ has the signature $(-,+)$.
The branes sets are $I_{1} = \{1,5\}$, $I_{2} = \{2,4,5\}$ and $I_{3} = \{3,4,5\}$.

We introduce the following set of  $\Gamma$-matrices
\begin{eqnarray}\label{10.4}
(\hat{\Gamma}^{A})= \begin{array}{ccccccccccc}
(\hat{\Gamma}^{a_{0}}_{(0)} &\otimes &1& \otimes &\textbf{1}_{2}&\otimes& \textbf{1}_{2} &\otimes &\textbf{1}_{2}&\otimes & \textbf{1}_{2},  \\
  \hat{\Gamma}_{(0)}& \otimes & 1 & \otimes & \hat{\Gamma}_{(2)} & \otimes & \hat{\Gamma}_{(3)} & \otimes & \hat{\Gamma}_{(4)} & \otimes & \hat{\Gamma}_{(5)} \\
   i\hat{\Gamma}_{(0)} & \otimes & 1& \otimes & \hat{\Gamma}^{a_{2}}_{(2)} & \otimes & \textbf{1}_{2} & \otimes & \textbf{1}_{2} & \otimes & \textbf{1}_{2}  \\
   \hat{\Gamma}_{(0)} & \otimes & 1 & \otimes & \hat{\Gamma}_{(2)} & \otimes & \hat{\Gamma}^{a_{3}}_{(3)} & \otimes & \textbf{1}_{2} & \otimes & \textbf{1}_{2}  \\
i\hat{\Gamma}_{(0)} & \otimes & 1& \otimes & \hat{\Gamma}_{(2)} & \otimes & \hat{\Gamma}_{(3)}& \otimes & \hat{\Gamma}^{a_{4}}_{(4)}& \otimes &\textbf{1}_{2} \\
  \hat{\Gamma}_{(0)}&  \otimes & 1 & \otimes & \hat{\Gamma}_{(2)} & \otimes & \hat{\Gamma}_{(3)} & \otimes & \hat{\Gamma}_{(4)} & \otimes & \hat{\Gamma}^{a_{5}}_{(5)}),
\end{array}
\end{eqnarray}
where $\hat{\Gamma}^{a_{i}}_{(i)}$ are $2\times 2$ $\Gamma$-matrices, $a_{i} = 1_{i},2_{i}$, $i=0,2,3,4,5$, corresponding to $M_{i}$, respectively, and the operators
 \begin{equation}\label{10.5}
\hat{\Gamma}_{(i)} = \hat{\Gamma}^{1_{i}}_{(i)}\hat{\Gamma}^{2_{i}}_{(i)}, \quad \hat{\Gamma}_{(5)} = \hat{\Gamma}^{1_{5}}_{(5)}\hat{\Gamma}^{2_{5}}_{(5)}
\end{equation}
satisfy
\begin{equation}\label{10.6}
(\hat{\Gamma}_{(i)})^{2}  = -\textbf{1}_{2},\quad (\hat{\Gamma}_{(5)})^{2} = \textbf{1}_{2},
\end{equation}
$i = 0,2,3,4$.

Consider $\eta$ in the form
$\eta = \eta_{0}(x)\otimes \eta_{1}(y_{1})\otimes\eta_{2}(y_{2})\otimes \eta_{3}(y_{3})\otimes\eta_{4}(y_{4})\otimes \eta_{5}(y_{5})$,
where $\eta_{i} = \eta_{i}(y_{i})$ is a 2-component spinor on $M_{i}$, $i=0,2,3,4,5$, $\eta_{1}(y_{1})$ is a 1-component spinor on $M_{1}$.

Due to (\ref{10.4}) and (\ref{10.6}) the operator
$\bar{D}_{m_{i}}$ acts on $\eta$ as
\begin{equation}\label{10.7}
\bar{D}_{m_{i}}\eta = \ldots\otimes\eta_{i-1}\otimes\left(D^{(i)}_{m_{i}}\eta_{i}\right)\otimes\eta_{i+1}\otimes \ldots,
\end{equation}
where $D^{(i)}_{m_{i}}$ is the spinorial covariant derivative corresponding to $M_{i}$, $i=0,2,3,4,5$, and $D^{(1)}_{m_{1}} = \partial_{m_{1}}$.
Thus the relations (\ref{2.13}) are satisfied if $\eta_{i}$  is a parallel spinor on  $M_{i}$, $i=0,1,2,3,4,5$, and $\eta_{1}(y_{1}) = \eta_{1}$
is constant.

The operator corresponding to the $M2$-brane is
\begin{equation}\label{10.8A}
\hat{\Gamma}_{[s]}= \hat{\Gamma}^{1_{1}}\hat{\Gamma}^{1_{5}}\hat{\Gamma}^{2_{5}} =
\hat{\Gamma}_{(0)}\otimes 1\otimes \hat{\Gamma}_{(2)}\otimes \hat{\Gamma}_{(3)}\otimes\hat{\Gamma}_{(4)}\otimes \textbf{1}_{2}
\end{equation}
for $s= I_{1}$, the operators corresponding to the $M5$-branes are
\begin{eqnarray}\label{10.8}
\hat{\Gamma}_{[s]} =\hat{\Gamma}^{1_{0}}\hat{\Gamma}^{2_{0}}\hat{\Gamma}^{1_{1}}\hat{\Gamma}^{1_{3}}\hat{\Gamma}^{2_{3}} =
\textbf{1}_{2}\otimes 1\otimes \hat{\Gamma}_{(2)}\otimes \textbf{1}_{2} \otimes \hat{\Gamma}_{(4)}\otimes \hat{\Gamma}_{(5)}
\end{eqnarray}
 for $s = I_{2}$ and
 \begin{eqnarray}\label{10.9}
\hat{\Gamma}_{[s]} =\hat{\Gamma}^{1_{0}}\hat{\Gamma}^{2_{0}}\hat{\Gamma}^{1_{1}}\hat{\Gamma}^{1_{2}}\hat{\Gamma}^{2_{2}} =
\textbf{1}_{2}\otimes 1 \otimes \textbf{1}_{2}\otimes \hat{\Gamma}_{(3)}\otimes \hat{\Gamma}_{(4)}\otimes \hat{\Gamma}_{(5)}
\end{eqnarray}
for $s = I_{3}$.

The projections (\ref{2.14}) are satisfied if
\begin{eqnarray}\label{10.10}
\hat{\Gamma}_{(j)}\eta_{j} = c_{(j)} \eta_{j}, \qquad c^{2}_{(j)} = -1,\nonumber \\
\hat{\Gamma}_{(5)}\eta_{5} = c_{(5)} \eta_{5}, \qquad c^{2}_{(5)} = 1,
\end{eqnarray}
$j=0,3,2,4$, and
\begin{eqnarray}\label{10.11}
c_{(0)}c_{(2)}c_{(3)}c_{(4)} = c_{1}, \quad
c_{(2)}c_{(4)}c_{(5)} = c_{2}, \quad c_{(3)}c_{(4)}c_{(5)} = c_{3}.
\end{eqnarray}

Eqs. (\ref{10.11}) have the following solutions
\begin{eqnarray}\label{10.12a}
c_{(0)} = ic_{1}c_{2}c_{3}\varepsilon_4, \quad
c_{(2)} =  -i c_{2} \varepsilon_4 \varepsilon_5, \\
c_{(3)} =  - i c_{3} \varepsilon_4 \varepsilon_5, \quad  c_{(4)} =  i \varepsilon_4, \quad
c_{(5)} = \varepsilon_{5},
\end{eqnarray}
where $\varepsilon_4 = \pm 1$, $\varepsilon_5 = \pm 1$.

For the field configuration (\ref{10.2}) and (\ref{10.3}) we obtain the following solution to SUSY equations
\begin{equation}\label{10.12}
\varepsilon = H^{-1/6}_{1} H^{-1/12}_{2} H^{-1/12}_{3} \eta_{0}(x) \otimes
\eta_{1}\otimes \eta_{2}(y_{2})\otimes\eta_{2}(y_{3})\otimes\eta_{4}(y_{4})\otimes \eta_{5}(y_{5}),
\end{equation}
where $\eta_{i}$, $i= 0 ,2,3,4,5$, are chiral parallel spinors defined on
$M_{i}$ $\left(D^{(i)}_{m_{i}}\eta_{i} = 0\right)$, obeying (\ref{10.10}) and (\ref{10.11}),  $\eta_{1}$ is constant.

Thus, the number of linear independent solutions is

\begin{eqnarray}\label{10.14}
N = 32 \mathcal{N} = \sum_{\begin{subarray}{l}\varepsilon_{4} = \pm 1, \\
\varepsilon_{5} = \pm 1 \end{subarray}}  n_{0}(i \varepsilon_4 c_{1} c_{2} c_{3})n_{2}(-i c_{2} \varepsilon_4 \varepsilon_5)  n_{3}(- i c_{3} \varepsilon_4 \varepsilon_5) n_{4}(i\varepsilon_4) n_{5}(\varepsilon_{5}),
 \end{eqnarray}
where $n_{j}(c_{j})$ is the number of chiral parallel spinors on $M_{j}$, $j=0,2,3,4,5$.

\textbf{Examples.}

Let $M_{0} = M_{2} = M_{3} = M_{4} = \mathbb{R}^{2}$ and $M_{1} = \mathbb{R}$.

\begin{description}
  \item[\{a\}]  Then for $M_{5} = \mathbb{R}^{1,1}$ one gets $\mathcal{N} = 1/8$.
  \item[\{b\}]  While in the case of  $M_{5} = \mathbb{R}^{1,1}_{*}/Z_{2}$, we find $\mathcal{N} = 1/16$ for any choice of brane sign factors.
\end{description}

\textbf{(ii)}

The second possible intersection of $M2$-brane and two $M5$-branes is defined on the manifold
 \begin{eqnarray}\label{11.1}
M_{0}\times M_{1}\times M_{2}\times M_{3}\times M_{4}\times M_{5}\times M_{6},
\end{eqnarray}
where $d_{0} = 3$, $d_{1}=d_{2}=d_{4}=d_{5}=d_{6}=1$, $d_{3}=3$.

The configuration is given in Fig. 7.

\begin{figure}[!h]
\begin{eqnarray}
g = \left\{
\begin{array}{ccccccccccccc}
- & - & - & - & - & - & - & - & \times & \times &\times & &:H_1 \nonumber \\
- & - & - & \times & - & \times & \times & \times & - & \times &\times & &:H_2  \\
- & - & \hbox to 0pt{\hss$\underbrace{\hskip1.6cm}_{M_0}$\hskip 1cm\hss} -&\underbrace{-}_{M_{1}} &\underbrace{\times}_{M_{2}} & \times & \times & \hbox to 0pt{\hss$\underbrace{\hskip1.6cm}_{M_3}$\hskip 1cm\hss} \times & \underbrace{\times}_{M_4} & \underbrace{-}_{M_5} & \underbrace{\times}_{M_6}  & &:H_3.
\end{array} \right.
\end{eqnarray}
 \caption{$M2\cap M5\cap M5$-intersection: the $M2$-brane intersect each of the two  $M5$-branes over a string, the $M5$-branes intersect over a 3-brane. $M_{0}$ is the totally transverse space,
 $M_{i}$,$i=1,2,3,4,5$, are the relatively transverse spaces, $M_{6}$ is the common worldvolume.}
\end{figure}

The solution describing the intersection of one electric brane and two magnetic ones  reads now
\begin{eqnarray}\label{11.2}
g = H^{1/3}_{1}H^{2/3}_{2}H^{2/3}_{3}\Bigl\{g^{0} + H^{-1}_{2}g^{1} + H^{-1}_{3}g^{2} + H^{-1}_{2}H^{-1}_{3}g^{3} + H^{-1}_{1}H^{-1}_{3}g^{4} + \nonumber\\
H^{-1}_{1}H^{-1}_{2}g^{5} +  H^{-1}_{1}H^{-1}_{2}H^{-1}_{3}g^{6} \Bigr\},\\ \label{11.2A}
F = c_{1}dH^{-1}_{1}\wedge \tau_{4}\wedge \tau_{5}\wedge \tau_{6} + c_{2}(\ast_{0}dH_{2})\wedge \tau_{2}\wedge \tau_{4} + c_{3}(\ast_{0} dH_{3})\wedge \tau_{1}\wedge \tau_{5},
\end{eqnarray}
where $c^{2}_{1} = c^{2}_{2} = c^{2}_{3} = 1$; $H_{1}$, $H_{2}$, $H_{3}$ are harmonic functions on $(M_{0},g^{0})$. The metrics $g^{i}$, $i=0,1,2,3,4,5$, have Euclidean signatures and we put the metric $g^{6} = -dt\otimes dt$.
The branes sets are $I_{1} = \{4,5,6\}$, $I_{2} = \{1,3,5,6\}$, $I_{3} = \{2,3,4,6\}$.

The corresponding set of gamma matrices matches with (\ref{9.3}) for two electric and one magnetic branes due to the same space
configuration. The expressions for $\hat{\Gamma}_{(i)}$, $(\hat{\Gamma}_{(i)})^{2}$, $i=0,3,$ coincide with (\ref{9.4}) and (\ref{9.5}) as well.

The 32-component spinor $\eta$ can be represented in the form
(\ref{9.5A})
$\eta = \eta_{0}(x)  \otimes \eta_{1} (y_{1}) \otimes
\eta_{2}(y_{2})  \otimes \eta_{3}(y_{3}) \otimes \eta_{4} (y_{4})
\otimes \eta_{5} \otimes \eta_{6}\otimes \chi$,
where $\eta_{i} = \eta_{i}(y_{i})$ is a 1-component spinor on
$M_{i}$, $i=1,2,4,5,6$, $\eta_{0} = \eta_{0}(x)$ is a 2-component
spinor on $M_{0}$ and $\eta_{3} = \eta_{3}(y_{3})$ is a
2-component spinor on $M_{3}$ and $\chi$
is an element of $V = \mathbb{C}^{2}\otimes \mathbb{C}^{2}\otimes
\mathbb{C}^{2}$.

For covariant derivatives we have  relations (\ref{9.6a}) and
(\ref{9.6}). Using the representation of gamma matrices (\ref{9.3}) the operators corresponding to $M$-branes are given by
\begin{equation}\label{11.6}
\hat{\Gamma}_{[s]}= \hat{\Gamma}^{1_{4}}\hat{\Gamma}^{1_{5}}\hat{\Gamma}^{1_{6}} = -\mathbf{1}_{2}\otimes 1 \otimes 1\otimes \mathbf{1}_{2}\otimes 1\otimes 1 \otimes 1 \otimes  B_{1}
\end{equation}
for  $s= I_{1}$,
\begin{equation}\label{11.6A}
\hat{\Gamma}_{[s]} =\hat{\Gamma}^{1_{0}}\hat{\Gamma}^{2_{0}}\hat{\Gamma}^{3_{0}}\hat{\Gamma}^{1_{2}}\hat{\Gamma}^{1_{4}} =  i\hat{\Gamma}_{0}\otimes 1 \otimes 1 \otimes \mathbf{1}_{2}\otimes 1 \otimes 1 \otimes 1\otimes B_{2}
\end{equation}
for $s = I_{2}$,
\begin{equation}\label{11.6B}
\hat{\Gamma}_{[s]} =\hat{\Gamma}^{1_{0}}\hat{\Gamma}^{2_{0}}\hat{\Gamma}^{3_{0}}\hat{\Gamma}^{1_{1}}\hat{\Gamma}^{1_{5}} = i\hat{\Gamma}_{(0)}\otimes 1 \otimes 1 \otimes \mathbf{1}_{2} \otimes 1  \otimes 1 \otimes 1\otimes B_{3}
\end{equation}
for  $s = I_{3}$.

Here
\begin{equation}\label{11.7}
B_{1} = \sigma_{2} \otimes \sigma_{2} \otimes \mathbf{1}_{2},
\quad B_{2} = \sigma_{3}\otimes \sigma_{1}\otimes \sigma_{3} ,
\quad B_{3} = \mathbf{1}_{2} \otimes \sigma_{2} \otimes \sigma_{1}
\end{equation}
are self-adjoint commuting operators.

As in the case with two $M2$-branes and one $M5$-brane we put
$(\hat{\Gamma}^{a_{0}}_{(0)}) = \left(\sigma_{1},\sigma_{2},
\sigma_{3}\right)$ and hence $\hat{\Gamma}_{(0)} =
i\textbf{1}_{2}$.

Due to the proposition from Appendix A there exists a basis of
eigenvectors
$\psi_{\varepsilon_{1},\varepsilon_{2},\varepsilon_{3}} \in V$
satisfying the relations
\begin{equation}\label{11.8}
B_{s}\psi_{\varepsilon_{1},\varepsilon_{2},\varepsilon_{3}} = \varepsilon_{s}\psi_{\varepsilon_{1},\varepsilon_{2},\varepsilon_{3}},
\end{equation}
where $\varepsilon_{1} =\pm 1$, $\varepsilon_{2} = \pm 1$, $\varepsilon_{3} = \pm 1$, $s =1,2,3$.

The solutions to eqs. (\ref{2.6}) corresponding to the field configuration from (\ref{11.2}), (\ref{11.2A}) are given rise by
\begin{equation}\label{11.9}
\varepsilon = H^{-1/6}_{1}H^{-1/12}_{2}H^{-1/12}_{3}\eta_{0}(x)\otimes\eta_{1}\otimes \eta_{2}\otimes \eta_{3}(y_{3})\otimes \eta_{4} \otimes \eta_{5}\otimes \eta_{6}\otimes \psi_{\varepsilon_{1},\varepsilon_{2},\varepsilon_{3}},
\end{equation}
where $\eta_{0}(x)$ and $\eta_{3}(y_{3})$ are parallel spinors on $M_{0}$ and $M_{3}$, $\eta_{i}$ are constant 1-dimensional spinors, $i=1,2,4,5,6$.

The parameters $\varepsilon_{s}$ obey the chirality restrictions
\begin{equation}\label{11.10}
- \varepsilon_{s} = c_{s},
\end{equation}
$s = 1,2,3$, following from relations (\ref{2.14}), (\ref{8.4c}) with $\chi = \psi_{\varepsilon_{1},\varepsilon_{2}, \varepsilon_{3}}$, (\ref{11.6})-(\ref{11.6B}), (\ref{11.8}).

The number of linear independent solutions given by (\ref{11.9})-(\ref{11.10}) can be computed as follows
\begin{equation}\label{11.7}
N = 32 \mathcal{N} = n_{0}n_{3},
\end{equation}
where $n_{0}$ and $n_{3}$ are numbers of parallel spinors on the manifolds $M_{0}$ and $M_{3}$, respectively.

\textbf{Example.}

Here the only example we have is the trivial one: $M_{0} = M_{2} =
\mathbb{R}^{3}$ with $\mathcal{N} = 1/8$ in agreement with
\cite{BREJS}.

\section{Conclusions}

In this paper  we have considered the generalized Killing
equations in $D=11$ supergravity  for triple  $M$-brane solutions
defined on the products of Ricci-flat manifolds.  The first
configuration with three electric branes $M2 \cap M2 \cap M2$ has
been studied earlier in \cite{IVD2}, while six others ones: $M2
\cap M2 \cap M5$, $M2 \cap M5 \cap M5$ (two configurations) and
$M5 \cap M5 \cap M5$ (three configurations) have been considered
here for the first time. Using the approach of \cite{IVD1,IVD2} we
have obtained explicit formulae for computing the amounts of
preserved supersymmetries for all triple $M$-brane configurations.
These formulae have generalized the relations obtained earlier by
several authors for  flat factor spaces $\mathbb{R}^{k_i}$
\cite{Ts1,BREJS}. The deduced fractional numbers of preserved SUSY
$\mathcal{N}$ depend upon certain numbers of (chiral or all)
parallel  spinors on some factor spaces and in several cases upon
brane sign factors $c_{s}$.

We have presented  examples of partially supersymmetric
configurations which do not belong to  the classification of
Bergshoeff et al. \cite{BREJS}. These examples use the following
factor spaces: $K3$ \cite{IVD2}, $\mathbb{C}^{2}_{*}/Z_2$ (for $M2 \cap M2
\cap M2$), $\mathbb{R}^{1,1}_{*}/Z_2$ (for case (i) of $M2 \cap M5 \cap
M5$ and case (ii) of $M5 \cap M5 \cap M5$) and $(\mathbb{R}^{1,1}_{*}/Z_2)
\times \mathbb{R}$ (for case (iii) of $M5 \cap M5 \cap M5$), a 4-dimensional
$pp$-wave manifold from \cite{FOF} and $(\mathbb{R}^{1,1}_{*}/Z_2) \times
\mathbb{R}^2$ (for case (i) of $M5 \cap M5 \cap M5$). This list of factor
spaces contains only few 4-dimensional Ricci-flat factor spaces
($K3$, $pp$-wave) which are not flat. All other two- and
three-dimensional factor spaces are flat. (Any $d =2,3$ Ricci-flat
space is flat.)

In three cases: $M2 \cap M2 \cap M2$ and $M5 \cap M5 \cap M5$ ((i)
and (ii)) we have presented examples where $\mathcal{N}$ depend upon
brane sign factors $c_s = \pm 1$.

An open problem here is to analyze  special solutions with certain
''near-horizon''  harmonic functions $H_{s}$ for which the
unbroken numbers of supersymmetries  might  be larger  then the
numbers $\mathcal{N}$ obtained here for generic $H_s$-functions.
In this case one should deal with Freund-Rubin-type solutions with
composite $M$-branes, see \cite{I-2} and references therein. Such
partially supersymmetric solutions will lead to certain relations
which contain  numbers of (chiral) Killing spinors on certain
Einstein factor spaces. This may be of interest in a context of
the AdS/CFT approach, its generalizations and applications
\cite{MaM}.

It is also a straightforward task to use the obtained  results for
studying  partially supersymmetric solutions (with Ricci-flat or
non-trivial flat factor spaces) in $IIA$- , $IIB$- and other ($d < 10$)
supergravitational models using dimensional reductions and
duality transformations. Another problem of interest may be
related to a search of ''pseudo-supersymmetric'' brane solutions
\cite{LuW} defined on a product of Ricci-flat manifolds by using  a
possible generalization of the approach from \cite{IVD1,IVD2}.

\section*{Acknowledgments}
The authors are grateful to D.V. Alekseevsky and H. Baum for helpful comments
on related issues. A.G. was supported in part by The Ministry of education and
science of Russian Federation, project 14.B37.21.2035.

 \renewcommand{\theequation}{\Alph{subsection}.\arabic{equation}}
 \renewcommand{\thesection}{}
 \renewcommand{\thesubsection}{\Alph{subsection}}
 \setcounter{section}{0}
 \appendix{ }

\section{Appendix}

Here we outline a proposition on simultaneous diagonalization of a set of linear idempotent operators
arising in the decomposition of gamma matrices for some cases of product spaces.
This proposition is a special case of the so-called ''$2^{-k}$-splitting'' theorem from \cite{IVD1}.

{\bf Proposition.}

{\em  Let  $B_{1}, \ldots, B_{k}: V \to V$ be a set of linear operators defined on the vector space
$V = \mathbb{C}^{2}\otimes...\otimes  \mathbb{C}^{2}$ (k-times) which are idempotent

 \begin{equation}\label{A.0}
  B^{2}_{i} = \mathbf{1}_{V}
  \end{equation}

and commute with each other

 \begin{equation}\label{A.1}
 B_{i} B_{j} = B_{j} B_{i},
 \end{equation}

$i,j = 1, ..., k$.

Let $A_{1}, \ldots A_{k}: V \rightarrow V$ be bijective operators  obeying the following relations

 \begin{equation}\label{A.2}
 A_{i}B_{i} + B_{i}A_{i} = 0,
 \end{equation}

 \begin{equation}\label{A.3}
 A_{i}B_{j} = B_{j}A_{i}, \quad i\neq j,
\end{equation}

$i,j = 1, ..., k$.

Then there exists a basis of $2^{k}$ vectors in $V$,
$\psi_{\varepsilon_{1},... ,\varepsilon_{k}}$,
$\varepsilon_{1},... ,\varepsilon_{k} = \pm 1$,
 which are eigenvectors of $B_i$:

\begin{equation}\label{A.4}
B_{i} \psi_{\varepsilon_{1},... ,\varepsilon_{k}} =
\varepsilon_{i} \psi_{\varepsilon_{1},...,\varepsilon_{k}},
\end{equation}
$i = 1, ..., k$. }

According to the proof of the theorem in \cite{IVD1}, the operator
$A_i$ defines an isomorphism between vector eigen-spaces $V_{...,
\varepsilon_{i}, ... }$ and $V_{..., - \varepsilon_{i}, ... }$
(all other indices $\varepsilon_{j}$, $j \neq i$, are coinciding).
Hence the basis may be found as follows. First we find a non-zero
``ground-state'' vector $\psi_{-1,... , -1}= \psi$ satisfying
\begin{equation}\label{A.5}
B_{i} \psi = - \psi
\end{equation}
for all $i$ and afterwards we put

\begin{equation}\label{A.7}
\psi_{\varepsilon_{1}, \ldots, \varepsilon_{k}}  \equiv
A^{(1+\varepsilon_{1})/2}_{1}\ldots A^{(1+\varepsilon_{k})/2}_{k}
\psi,
\end{equation}
where $\varepsilon_{1},\ldots, \varepsilon_{k} = \pm 1$. Here
$A^{0}_{i} \equiv \mathbf{1}_{V}$ is identity operator on $V$.

Let us consider three examples of the operators $B_1, B_2, B_3$,
which appeared in Section 3 (in the cases of intersections $M5\cap M5 \cap M5$ (\textbf{iii}), $M2\cap M2\cap M5$ and $M2\cap M5\cap M5$ (\textbf{ii})).

{\bf Example 1.}

Let
\begin{equation}\label{A.8}
B_{1} = -\sigma_{1}\otimes \sigma_{2}\otimes \mathbf{1}_{2}, \quad
B_{2} = \sigma_{3}\otimes\sigma_{1}\otimes\sigma_{3}, \quad B_{3}
= - \mathbf{1}_{2}\otimes\sigma_{2}\otimes \sigma_{1}.
\end{equation}

A set of the operators $A_{s}$, $s=1,2,3$, obeying
(\ref{A.2})-(\ref{A.3}) may be chosen as

\begin{equation}\label{A.8a}
 A_{1} = \sigma_{3}\otimes \mathbf{1}_{2}\otimes\mathbf{1}_{2},
 \quad A_{2} = \sigma_{1}\otimes\mathbf{1}_{2}\otimes
 \mathbf{1}_{2},\quad A_{3} =
 \mathbf{1}_{2}\otimes\mathbf{1}_{2}\otimes \sigma_{3}.
 \end{equation}


{\bf Example 2.}

For
\begin{equation}\label{A.9}
B_{1} = \sigma_{1}\otimes \textbf{1}_{2}\otimes \sigma_{3}, \quad
B_{2} = \sigma_{2}\otimes\sigma_{3}\otimes \sigma_{1}, \quad B_{3}
= \mathbf{1}_{2}\otimes \sigma_{3}\otimes \mathbf{1}_{2},
\end{equation}
the operators $A_{s}$  read
\begin{equation}\label{A.9a}
A_{1} = \sigma_{2}\otimes \mathbf{1}_{2}\otimes \sigma_{3}, \quad
A_{2} = \sigma_{1} \otimes \mathbf{1}_{2} \otimes \mathbf{1}_{2},
\quad A_{3} =  \sigma_{1} \otimes \sigma_{1}\otimes
\mathbf{1}_{2}.
\end{equation}


{\bf Example 3.}

Let
\begin{equation}\label{A.10}
B_{1} = \sigma_{2}\otimes \sigma_{2}\otimes\mathbf{1}_{2}, \quad
B_{2} = \sigma_{3}\otimes\sigma_{2}\otimes\sigma_{3}, \quad B_{3}
= \mathbf{1}_{2}\otimes\sigma_{2}\otimes\sigma_{1}.
\end{equation}

Then $A_{s}$-operators can be presented in the following form
\begin{equation}\label{A.10a}
A_{1} = \sigma_{3}\otimes \mathbf{1}_{2}\otimes \mathbf{1}_{2},
\quad A_{2} =
\sigma_{2}\otimes\mathbf{1}_{2}\otimes\mathbf{1}_{2}, \quad A_{3}
= \mathbf{1}_{2}\otimes \mathbf{1}_{2}\otimes \sigma_{3}.
\end{equation}


\end{document}